%% file: main.tex
\documentclass[11pt]{article}

\usepackage[T1]{fontenc}
\usepackage{amsmath}
\usepackage{newtxtext,newtxmath}
\usepackage[letterpaper,top=0.88in,bottom=0.90in,
            left=1in,right=1in,headheight=22pt]{geometry}
\usepackage{booktabs}
\usepackage{graphicx}
\usepackage{xcolor}
\usepackage{enumitem}
\usepackage{microtype}
\usepackage{setspace}
\usepackage{abstract}
\usepackage{titlesec}
\usepackage{fancyhdr}
\usepackage{caption}
\usepackage{xurl}
\usepackage{etoolbox}
\usepackage{placeins}
\definecolor{papergray}{HTML}{555555}
\usepackage[authoryear,round]{natbib}
\setcitestyle{aysep={,},yysep={;},notesep={; }}
\usepackage[hidelinks]{hyperref}
\usepackage{tikz}
\usetikzlibrary{arrows.meta,positioning,shapes.geometric,calc,decorations.pathreplacing,fit,backgrounds}
\usepackage{pgfplots}
\pgfplotsset{compat=1.18}

\hypersetup{
  pdftitle={From Word Counts to Context: Topic Models for Asset Pricing},
  pdfauthor={Kevin Foley, Jonathan Hartadi, Shivesh Prakash, and Swapnil Vatsal},
  pdfsubject={Narrative asset pricing and contextual topic models},
  pdfkeywords={narrative asset pricing, topic models, sentence transformers, financial news, IPCA}
}

\setlist{nosep,leftmargin=1.5em}

\titleformat{\section}[hang]
  {\Large\bfseries\color{black}}{\thesection}{0.7em}{}
\titleformat{\subsection}[hang]
  {\large\bfseries\color{black}}{\thesubsection}{0.65em}{}
\titleformat{\subsubsection}[hang]
  {\normalsize\bfseries\color{black}}{\thesubsubsection}{0.6em}{}
\titlespacing*{\section}{0pt}{2.2ex plus 0.6ex minus 0.2ex}{1.0ex}
\titlespacing*{\subsection}{0pt}{1.8ex plus 0.4ex minus 0.2ex}{0.7ex}

\fancypagestyle{paperstyle}{%
  \fancyhf{}
  \fancyhead[L]{\footnotesize\color{papergray}\textsc{From Word Counts to Context}}
  \fancyhead[R]{\footnotesize\color{papergray}\thepage}
  \renewcommand{\headrulewidth}{0.35pt}
  \renewcommand{\headrule}{\hbox to\headwidth{\color{bordergray}\leaders\hrule height \headrulewidth\hfill}}
  
}
\AtBeginEnvironment{thebibliography}{\small\setlength{\bibsep}{0.45em}}

\definecolor{navyblue}{RGB}{20, 55, 90}
\definecolor{midblue}{RGB}{40, 90, 135}
\definecolor{softblue}{RGB}{236, 243, 250}
\definecolor{accentteal}{RGB}{30, 120, 130}
\definecolor{softteal}{RGB}{235, 246, 247}
\definecolor{accentorange}{RGB}{190, 80, 25}
\definecolor{softorange}{RGB}{253, 244, 238}
\definecolor{darkgray}{RGB}{40, 45, 50}
\definecolor{midgray}{RGB}{100, 110, 120}
\definecolor{lightgray}{RGB}{246, 248, 250}
\definecolor{cardbg}{RGB}{255, 255, 255}
\definecolor{bordergray}{RGB}{205, 205, 205}

\newcommand{\orcidicon}{%
  \href{https://orcid.org/0009-0007-4120-0921}{%
    \raisebox{-0.8pt}{\includegraphics[height=9pt]{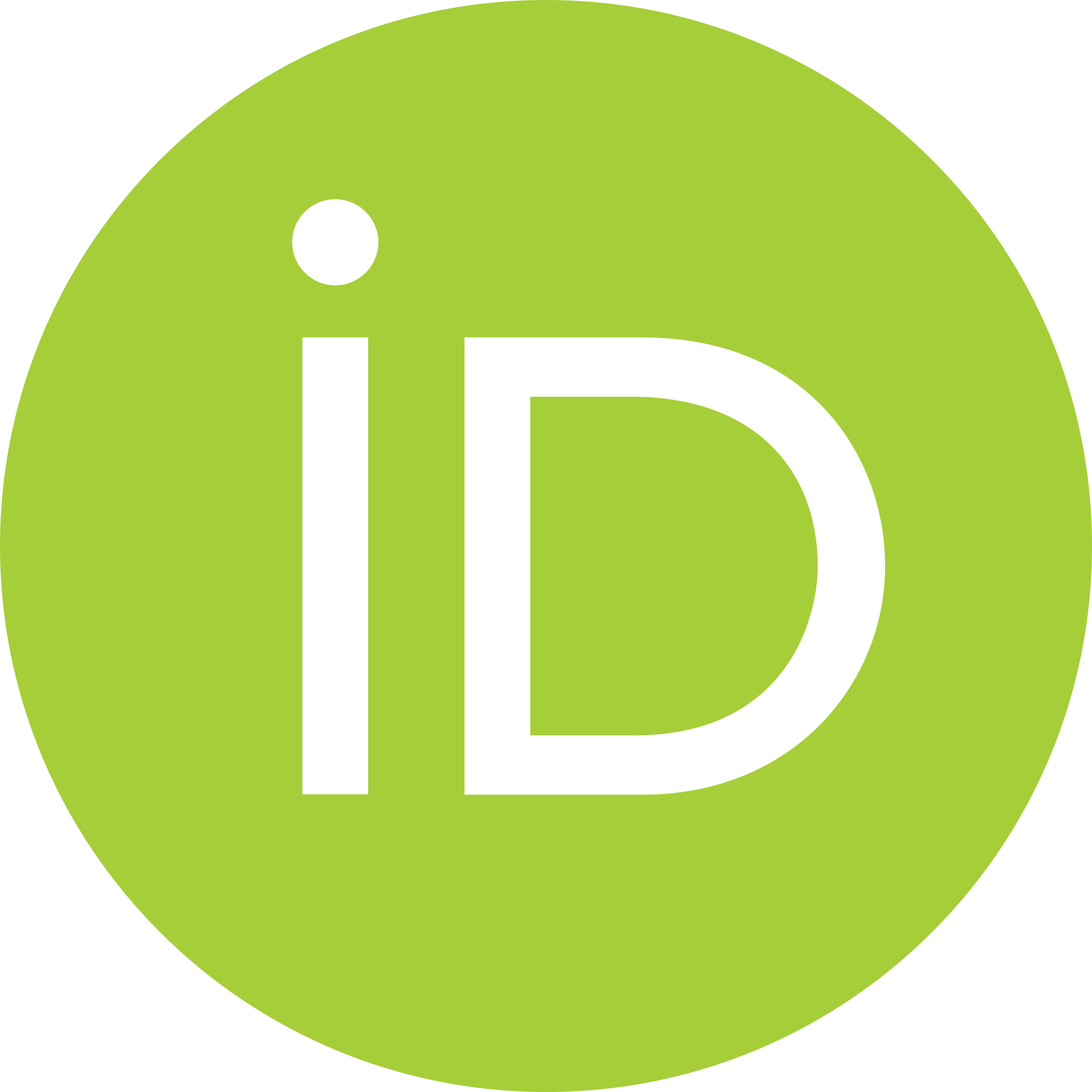}}}}

\renewcommand{\maketitle}{%
  \thispagestyle{empty}
  \vspace*{0.42in}
  \begin{center}
    {\fontsize{25}{30}\selectfont\bfseries
      From Word Counts to Context:\\[0.12em]
      Topic Models for Asset Pricing\par}
    \vspace{1.35em}
    {\large
      Kevin Foley\textsuperscript{*}\quad
      Jonathan Hartadi\textsuperscript{*}\\[0.35em]
      Shivesh Prakash\textsuperscript{*,\(\dagger\)}\,\orcidicon\quad
      Swapnil Vatsal\textsuperscript{*}\par}
    \vspace{0.75em}
    {\normalsize University of California, Berkeley\par}
    {\small Berkeley, California, USA\par}
    \vspace{0.85em}
    {\small
      \href{mailto:kevinfoley3@berkeley.edu}{kevinfoley3@berkeley.edu}
      \enspace\textperiodcentered\enspace
      \href{mailto:johartadi23@berkeley.edu}{johartadi23@berkeley.edu}\\[0.25em]
      \href{mailto:shivesh_prakash@berkeley.edu}{shivesh\_prakash@berkeley.edu}
      \enspace\textperiodcentered\enspace
      \href{mailto:swapnil_vatsal@berkeley.edu}{swapnil\_vatsal@berkeley.edu}\par}
    \vspace{0.65em}
    {\small\color{papergray}
      \href{https://shivesh777.github.io/bert_lda/}{%
        Project page:\enspace shivesh777.github.io/bert\_lda}\par}
  \end{center}
  \vspace{1.45em}
  \begingroup
    \renewcommand{\thefootnote}{\fnsymbol{footnote}}%
    \footnotetext[1]{Equal contribution; authors listed alphabetically by last name.}%
    \footnotetext[2]{Corresponding author.}%
  \endgroup
}

\begin{document}
\maketitle
% ---------------------------------------------------------------------
\begin{abstract}

News may reveal systematic risk, but whether its context enhances the construction of systematic risk factors is still unclear. We seek to test whether utilizing a sentence transformer represents an improvement over techniques such as Latent Dirichlet Allocation (LDA) in the coherence of topic term lists generated from unstructured text data. To test this, the same collection of unstructured text data comprising of 394,661 articles and the same downstream financial portfolio construction pipeline were applied with the text layer differing, including the length of article text each model used and how topic terms were ranked: we benchmark LDA against a frozen sentence transformer with k-means clustering. We find that the sentence transformer branch had higher observed scores both in terms of coherence (measured by NPMI) as well as financial performance (measured by Sharpe), although the available tests do not establish outperformance. Further exploratory specifications such as utilizing spherical clustering and multi-horizon exposures had an observed excess-return Sharpe of 1.03 for the combined model. We believe that there is some promise in applying context-aware techniques on unstructured news text, but stricter tests using only information available at each date and broader datasets may be required to enhance the confidence in the observed performance.

\end{abstract}

\vspace{0.8em}
\begin{center}
\small
\textbf{Keywords:}
narrative asset pricing; topic models; sentence transformers;
financial news; Sparse IPCA
\end{center}
\vfill
\clearpage
\pagestyle{paperstyle}

% ---------------------------------------------------------------------
\section{Introduction}
\label{sec:intro}

Systematic risk can be reflected in the language of financial news \citep{bybee2023narrative}. New technologies that can quantitatively interpret large amounts of unstructured text have created new opportunities to construct signals from language and test whether they contain information about risk and returns \citep{bybee2023narrative,kakhbod2026fedmind,kakhbod2025creative,kakhbod2025finfluencers,kakhbod2025nolbert}. Financial news is a natural setting for this question because large volumes of market-relevant text are published every day \citep{dong2024fnspid}.

Traditional topic models such as Latent Dirichlet Allocation (LDA) identify recurring themes by modeling patterns in word counts across documents \citep{blei2003lda}. As a bag-of-words model, however, LDA treats words as exchangeable and therefore does not directly represent word order or local context. Transformer-based language models instead provide contextual representations in which passages with similar meanings can be placed near one another even when their exact word usage differs \citep{reimers2019sentencebert}. We therefore ask whether replacing the LDA text layer with a frozen sentence transformer produces more coherent financial-news narratives and changes downstream asset-pricing performance while leaving the subsequent financial methodology fixed.

Our analysis builds on the narrative asset-pricing framework of \citet{bybee2023narrative}. Their approach converts news into daily narrative attention, measures unexpected changes in that attention, estimates stock-level covariances with the resulting narrative shocks, and uses Sparse IPCA to extract a small number of latent systematic risk factors. Intuitively, a shock to attention around oil supply may affect oil producers and airlines differently, creating a cross-section of stock exposures to that narrative. The estimated risk factors, however, are latent combinations of many such narrative exposures rather than factors corresponding one-for-one to individual topics.

Our central comparison changes the text representation while holding the downstream pipeline fixed. Both LDA and the transformer are estimated on the same FNSPID financial-news corpus \citep{dong2024fnspid}, produce $L=60$ topics, and feed daily narrative attention into the same shock construction, covariance characteristics, Sparse IPCA estimation, and out-of-sample portfolio evaluation. We evaluate the text layer using topic coherence, measured by Normalized Pointwise Mutual Information (NPMI), and the financial model using out-of-sample portfolio performance. One important limitation of the comparison is that LDA observes the full cleaned article, whereas the sentence transformer uses up to three lead passages, so the change in representation is not perfectly isolated from the amount of text supplied to each model. We therefore interpret the comparison accordingly.

\noindent\textbf{Contributions.}
Our contributions are threefold. First, we provide a controlled comparison of the narrative asset-pricing pipeline in which the downstream construction of shocks, stock-level covariance characteristics, Sparse IPCA factors, and out-of-sample portfolios is held fixed while the text layer changes. Second, we replace the LDA representation with a frozen sentence transformer and evaluate whether contextual embeddings produce more coherent narrative topics and different downstream pricing performance on the same financial-news corpus. Third, we study two exploratory extensions (spherical clustering of normalized embeddings and multi-horizon stock--narrative covariances) and report their incremental effects separately from the controlled LDA-versus-transformer comparison.

Overall, the analysis isolates how much the representation of financial news can matter once narrative signals are passed through a common asset-pricing framework.

\begin{figure}[!t]
\centering
\input{figures/fig1_pipeline.tex}
\caption{End-to-end narrative asset-pricing architecture. Financial news text is mapped into contextual semantic embeddings and clustered into $L=60$ narrative topics on the unit hypersphere. Unexpected deviations from trailing 5-day attention form narrative shocks, whose rolling stock-level covariances provide characteristic instruments for Sparse IPCA. The model extracts $K=3$ latent narrative risk factors that price the cross-section and form an OOS MVE tangency portfolio.}
\label{fig:pipeline_overview}
\end{figure}
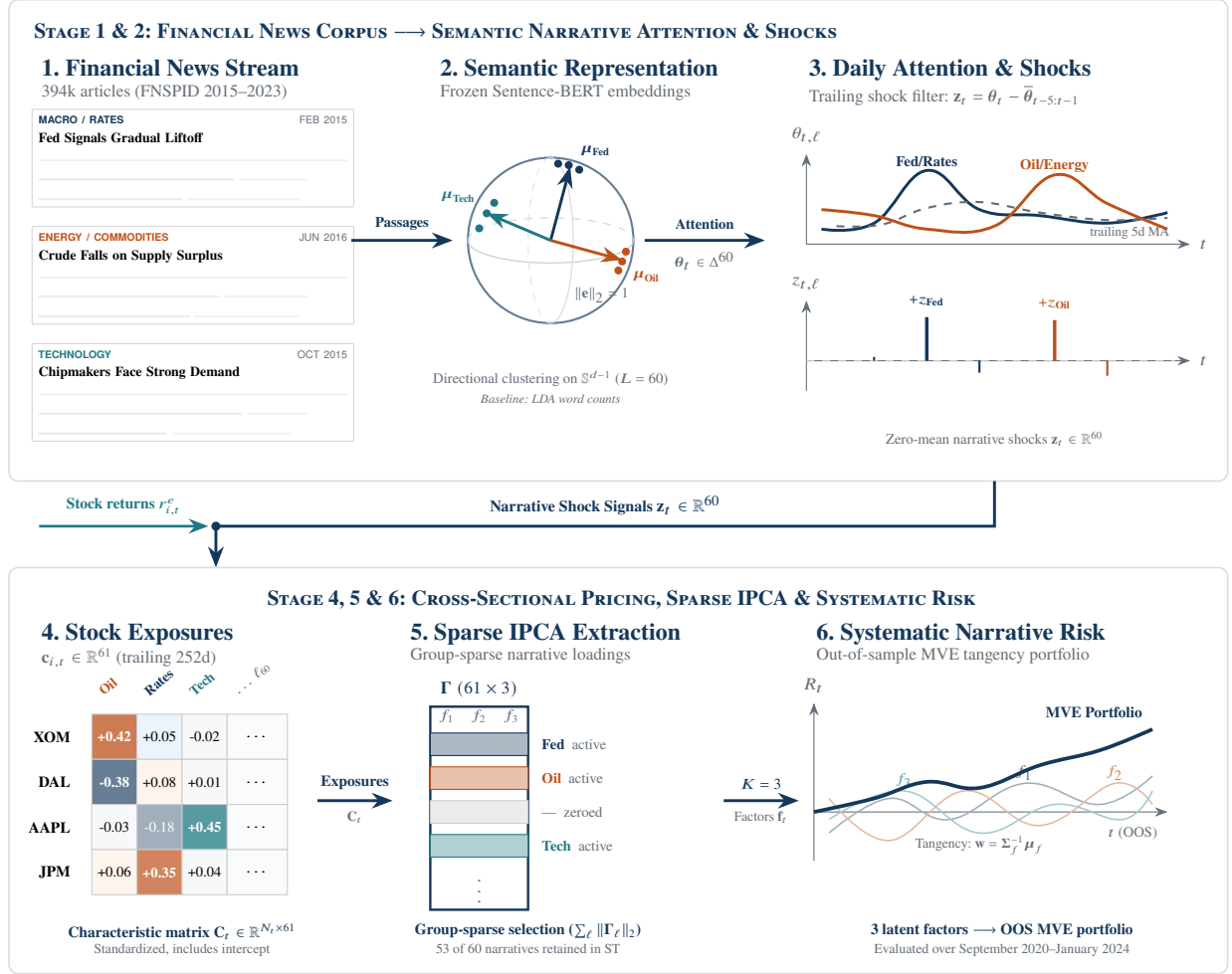
\FloatBarrier

% ---------------------------------------------------------------------
\section{Literature Review}
\label{sec:related}

Work that turns documents into numbers and then relates those numbers to
an economic outcome spans several literatures, and the survey of
\citet{gentzkow2019text} organizes it around that two-step structure: a
choice of text representation followed by a choice of downstream model.
We group prior work the same way and proceed by method, earliest first.
Section~\ref{sec:related:classical} covers count-based topic models,
which supply our baseline text layer and the pipeline we replicate.
Section~\ref{sec:related:transformer} covers contextual and
language-model-based topic models, which supply our alternative text
layer. Section~\ref{sec:related:nonnews} covers the same tools applied
to financial text other than news, where the corpus rather than the
method is what differs from our setting.
Section~\ref{sec:related:positioning} then states which of these results
can be placed beside ours and which cannot.

\subsection{Non-Transformer Topic Models}
\label{sec:related:classical}

Latent Dirichlet Allocation represents each document as a mixture over
topics and each topic as a distribution over vocabulary terms, with
Dirichlet priors on both \citep{blei2003lda}. Because the likelihood
depends on a document only through its word counts, words are
exchangeable and word order carries no information, which is the
assumption that Section~\ref{sec:related:transformer} relaxes. LDA is
our baseline text layer, fitted at the same topic count as the
transformer branch so that the two remain comparable.

Estimation is usually by collapsed Gibbs sampling, which repeatedly
reassigns individual term occurrences to topics and recovers the topic
and document distributions from the resulting assignment counts
\citep{griffiths2004finding}. This is the procedure our implementation
runs, and it is also the reason a single reported fit is one draw from a
posterior rather than a unique solution, a caveat we return to in
Section~\ref{sec:discussion}.

Applied to news, LDA has been used to build a low-dimensional summary of
what the business press is discussing. \citet{bybee2024business} fit 180
topics to the front section of the Wall Street Journal from 1984 to 2017
and show that the resulting attention series track macroeconomic
conditions and forecast business-cycle variables.
\citet{bybee2023narrative} carry the same topic estimates into asset
pricing, converting attention into shocks, shocks into stock-level
covariance characteristics, and those characteristics into a small set
of latent risk factors. We adopt that second pipeline without change
from the shock construction onward, and differ in the corpus, the topic
count, and the text layer itself.

A separate strand lets returns inform the text model directly.
\citet{ke2019predicting} screen for words whose frequency covaries with
subsequent returns and estimate a sentiment score from them, which makes
the text representation supervised by the outcome it is later used to
predict. We deliberately avoid this in both branches, because the
comparison we are making requires that neither text layer observe
returns.

Closest in spirit to our own construction is work that abandons the
generative likelihood and clusters distributed representations instead.
\citet{cong2025textual} build word embeddings, group them into
interpretable clusters, and use the resulting loadings as textual
factors for asset-pricing and prediction tasks. That design shares our
two-step structure of clustering followed by term-based interpretation,
and differs in that it clusters word-level vectors rather than
contextual representations of passages, which is the step taken by the
methods in Section~\ref{sec:related:transformer}.

None of these results is a numerical benchmark for ours. Published topic
counts, coherence values, and forecasting statistics are specific to
their corpora and tasks, so the only like-for-like comparison available
to us is between our own two branches, which share a corpus, a
vocabulary, and a topic count by construction.

\subsection{Transformer- and LLM-Based Topic Models}
\label{sec:related:transformer}

A second group of methods replaces the word-count representation with a
learned one. Transformer encoders trained with a masked language
objective produce word representations that depend on the surrounding
text, so the same word receives different representations in different
sentences \citep{devlin2019bert}. Sentence-BERT adapts such an encoder
so that an entire passage is summarized by a single vector and two
passages can be compared by the cosine of the angle between their
vectors \citep{reimers2019sentencebert}. Neither paper is about topic
modeling, but together they supply the representation used by the
methods below and by our own text layer.

Embedded topic models keep a generative structure while placing words
and topics in a common vector space, so a topic is a direction in that
space rather than an unstructured list of word probabilities
\citep{dieng2020etm}. This retains the probabilistic reading that LDA
offers and handles large vocabularies better, but each document is
still represented by its word counts rather than by a contextual
encoder. We differ in that our article weights come from an encoder
applied to running text, and we keep a distribution over terms only to
describe the resulting topics.

A closer precedent treats topic discovery as clustering in a vector
space. \citet{sia2020tired} show that clustering pretrained word
embeddings, with a weighting and reranking step applied to the
resulting word lists, produces topics competitive with LDA at a small
fraction of the estimation cost. Top2Vec clusters jointly embedded
document and word vectors and reads each topic's terms off the words
closest to its cluster center \citep{angelov2020top2vec}. Our text
layer belongs to this group, with two differences. We cluster
contextual representations of article passages rather than static word
vectors, and we score terms against a vocabulary fixed in advance, so
that our topics can be compared term for term with the LDA topics of
Section~\ref{sec:methods:lda}.

Two later methods are closer still. Contextualized topic models feed a
pretrained document embedding into a neural topic model and report
higher coherence than count-based inputs \citep{bianchi2021pretraining}.
BERTopic combines a transformer encoder, a dimensionality reduction
step, and density-based clustering with a class-based TF-IDF procedure
that gives each cluster a readable list of terms
\citep{grootendorst2022bertopic}. We follow BERTopic in separating the
clustering step from the term-scoring step, and we use its class-based
TF-IDF construction, but we assign passages softly rather than to a
single cluster, and we do not reduce the embedding dimension before
clustering. The number of topics is therefore fixed at $L$ by
construction rather than chosen by a density criterion, which matters
here because the comparison with LDA requires both models to produce
the same number of topics.

The most recent work prompts a large language model to propose topics
and assign documents to them, which produces labels that human readers
judge more interpretable than those of count-based models
\citep{pham2024topicgpt}. We do not take this route. A generative
labeler was itself trained on text of uncertain vintage, so the
assignments it produces may reflect information that postdates the
article being labeled. Related work in finance has responded by
building language models whose training data is restricted to text
available before a stated date \citep{kakhbod2025nolbert}. Our frozen
encoder does not satisfy that standard either, a point we return to in
Section~\ref{sec:discussion}, but a clustering of fixed embeddings at
least leaves the assignment rule transparent and independent of any
generated text. That independence matters since generated text can
differ across prompts: default responses overrepresent certain
demographics, and the bias changes when the prompt specifies a
different demographic \citep{fedyk2026ai}.

Finally, this literature is evaluated largely by automated coherence
measures, and that practice has been questioned.
\citet{hoyle2021broken} find that automated coherence agrees poorly
with human judgments of topic quality in exactly the regimes where
competing models are usually compared. We therefore report NPMI
\citep{roder2015coherence} as a descriptive summary of term lists
computed on a common vocabulary, and we do not treat it as evidence
that one text layer recovers more economically meaningful narratives
than the other.

\subsection{Topic Models on Non-News Financial Text}
\label{sec:related:nonnews}

Regulatory filings are the most heavily studied financial corpus.
\citet{dyer2017evolution} fit LDA to the full text of 10-K filings and
document how the composition of mandatory disclosure has shifted over
two decades, using topics as a descriptive summary rather than as an
input to a pricing model. Closer to our task,
\citet{lopezlira2019risk} applies a topic model to the risk-factor
sections of 10-Ks and forms return factors from the resulting
exposures, which makes it the nearest published analogue to what we do
with news. More broadly, advances in machine learning and AI have expanded the set of economically important firm characteristics that can be measured and studied, including dimensions of intangible capital, innovation, and organizational structure \citet{gentzkow2019text,chen2023process}.
The corpora differ in a way that matters for the downstream
design: filings are firm-authored, periodic, and regulated, so they
yield a firm-by-year panel rather than the daily market-wide attention
series that the shock construction requires.

Contextual encoders have entered this literature through the same door.
\citet{huang2023finbert} pretrain a BERT model on corporate filings,
analyst reports, and earnings calls, and show that it extracts
information that dictionary methods miss, particularly where meaning
depends on context. This supports the premise behind our transformer
branch, although it is a supervised classification result on firm
disclosure rather than an unsupervised topic model on news. More broadly, related work in financial economics uses interpretable machine-learning methods to extract economically meaningful structure from high-dimensional data in settings ranging from mortgage lending to equity portfolio choice and corporate bond returns \citet{bell2025fairness,bell2026alphaglass,bell2026glassbox}.

Beyond filings, the same tools have been applied to central bank
communication, patents, and social media.
\citet{kakhbod2026fedmind} extract latent dimensions from Federal
Reserve records, \citet{kakhbod2025creative} and
\citet{hochberg2026patents} estimate measures from patent text, and
\citet{kakhbod2025finfluencers} study retail-facing social media
content. These establish the breadth of text-based measurement in
finance, but the documents are produced by different authors under
different incentives and arrive on different schedules than daily
newswire copy.

We therefore treat this group as motivation rather than as a benchmark.
Neither the topic counts nor the coherence levels reported on these
corpora transfer to ours, and none of them produces the daily attention
series that the pipeline of \citet{bybee2023narrative} consumes.

\subsection{Positioning and Comparability}
\label{sec:related:positioning}

The comparison closest to ours is \citet{bybee2023narrative}, and even
that one is partial. We reuse their construction of shocks, covariance
characteristics, sparse estimation, and portfolio evaluation, so our LDA
branch is the same method applied to a different corpus over a different
period. Their Wall Street Journal sample runs from 1984 to 2017 and
supports 180 topics; ours is FNSPID newswire text from 2015 to 2023
\citep{dong2024fnspid} and supports 60. A difference in reported
performance between the two therefore confounds the corpus, the sample
period, and the topic count, and we do not read it as a replication
success or failure. \citet{bybee2024business} uses the same topics for
macroeconomic forecasting, which is a different task again.

The factor-model stage carries its own comparability problem. Sparse
IPCA estimates stock loadings as a function of observable
characteristics rather than from returns alone
\citep{kelly2019characteristics}, and the number of latent factors is a
modeling choice rather than an estimated quantity. Papers reporting more
or fewer factors are not thereby describing better or worse models, so
factor counts do not rank across studies.

Neither do Sharpe ratios. A strategy selected as the best among many
specifications has an upward-biased performance estimate, and the size
of that bias grows with the number of specifications tried and shrinks
with the length of the evaluation window \citep{bailey2014deflated}.
Our own evaluation window is 41 months and our specifications were
chosen after inspecting it, so the appropriate comparison for our
numbers is against benchmarks computed on the same months, not against
Sharpe ratios reported elsewhere.

Coherence values are equally corpus-bound. NPMI is computed from term
co-occurrence within a particular collection, so its level reflects that
collection's vocabulary and document lengths as much as the model's
quality \citep{roder2015coherence}, and automated coherence tracks human
judgment poorly in the range where models are typically compared
\citep{hoyle2021broken}. The coherence figures reported by
\citet{grootendorst2022bertopic} and \citet{sia2020tired} on general
corpora are therefore not a scale against which ours can be read. Only
our within-paper comparison, which holds the corpus and vocabulary
fixed, is interpretable.

One caveat applies to every embedding-based method discussed above,
including ours. A pretrained encoder has seen
text from a period that overlaps the evaluation sample, so its
representations may encode information that was not available on the
date of the article being encoded. Work in this area has begun
restricting training data to text available before a stated date in
order to remove that channel \citep{kakhbod2025nolbert}. Our encoder
does not meet that standard, which bounds how strongly the transformer
results should be read, and we return to this in
Section~\ref{sec:discussion}.

% ---------------------------------------------------------------------

\section{Methods}
\label{sec:methods}

Figure~\ref{fig:pipeline_overview} provides the end-to-end schematic
for the methodology developed below, from financial news text to
narrative attention shocks, stock covariance exposures, latent
systematic risk factors, and out-of-sample portfolio allocations.

\subsection{Data and Preprocessing}
\label{sec:methods:data}

The dataset we used consists of 394,661 financial news articles from FNSPID \citep{dong2024fnspid}, covering January 2015 through December 2023. Both models use the same article sample. During preprocessing and collection, we retained only the articles with valid dates, stock tickers, and article text. Articles that merely included a summary or were less than 500 characters were excluded. Additionally, we identified duplicate articles using their URLs or matching dates and titles and removed those articles from the sample. Article bodies were capped at 4,000 characters and combined with their headlines. A document index keeps the text, publication dates, and model outputs aligned.

For the LDA topic model \citep{blei2003lda}, we converted the retained text in the corpus into a matrix of term counts. We removed things such as URLs and nonalphabetic characters, converted text into lowercase, and normalized spacing. Next, we constructed a vocabulary list and excluded common stopwords such as ``shares,'' ``company,'' and ``Reuters.'' Words or phrases not included in the vocabulary selection were further purged from the corpus. Furthermore, candidate terms for our list had to appear in at least ten articles. Our final vocabulary list consisted of 15,000 terms: 7,467 words and 7,533 word pairs. This vocabulary was then fixed and used to count terms across the entire corpus. Appendix Exhibit~\ref{tab:preprocessing} summarizes the preprocessing steps for both models.

The preprocessing for the transformer differed from that used for LDA. The transformer received the original wording of the retained text to preserve word ordering and surrounding context. We divided each article into chunks of up to 200 words, excluded chunks with fewer than eight terms, and retained up to three opening chunks. These chunks were encoded with the pretrained MiniLM sentence transformer \citep{reimers2019sentencebert}, and the embeddings were normalized to unit length. The encoder remained frozen throughout our analysis. Both models produced 60 topics and shared the same vocabulary list. The shared vocabulary list supplies the LDA's input and the transformer's topic descriptions and coherence evaluation, but it does not constrain the transformer's tokenizer.

We aggregated article topic weights into daily attention ``shocks'' using each article's term count. We defined attention shocks as deviations of daily topic attention from its average over the previous five days \citep{bybee2023narrative}. Additionally, attention ``shocks'' recorded on weekends or holidays are carried forward to the next trading day. Both the LDA and transformer use the same CRSP return panel \citep{crspstockdatabase}, with monthly stock returns compounded from daily observations; we required at least 15 valid observations per month. The resulting financial sample contains 95,071 stock--month observations from August 2015 through January 2024.
The first 61 months form the initial financial estimation sample, followed by 41 evaluation months from September 2020 through January 2024. Financial models are subsequently refitted on expanding samples. However, the vocabulary list and topic estimates use text from the full news sample. The reported out-of-sample results therefore refer to the financial estimation stage rather than a fully chronological evaluation of the text and return models together.

\subsection{Baseline Text Layer: LDA}
\label{sec:methods:lda}

We fit our LDA baseline using the article counts prepared in Section~\ref{sec:methods:data}. We use 60 topics in total, the same number as in the transformer model, to hold the topic count fixed across the comparison. An article is a mixture of several topics, with the weight on each topic determined by the number of terms drawn from that topic in the article \citep{blei2003lda}.

We write the counts for article $d$ as $w_d$, a vector with $V=15{,}000$ entries, and their sum as $N_d$. Under LDA,

\begin{equation}
    w_d \mid \theta_d,\Phi
    \sim \operatorname{Multinomial}\!\left(N_d,\Phi\theta_d\right).
    \label{eq:lda}
\end{equation}

\noindent$\Phi$ has one column for each of the $L=60$ topics. A column $\phi_\ell$ is a probability distribution over words and phrases from the vocabulary defined in Section~\ref{sec:methods:data}, so its entries are nonnegative and sum to one. We describe each topic using its most probable terms. There are no topic names supplied to the model beforehand.

The article weights are collected in $\theta_d$. These also sum to one, but unlike the columns of $\Phi$, they differ across articles. Consider two news articles that both discuss employment. One might focus on layoffs and slowing demand, while the other describes hiring and business expansion. Their use of terms can lead to different topic mixtures even though they concern the same broad subject. The model uses the counts of the retained vocabulary terms and, unlike the transformer model, does not account for their position in the text.

We place Dirichlet priors on the topic distributions and article weights and estimate the model by collapsed Gibbs sampling \citep{griffiths2004finding}. Our implementation uses tomotopy with 800 iterations and a random seed of zero. Topic assignments for term occurrences are repeatedly updated during sampling, which eventually leads to the estimates of $\Phi$ and $\theta_d$ in our model. We then average the article weights within each day, weighted by term counts, to obtain the attention series used in the \citet{bybee2023narrative} pipeline. This fit uses the full news corpus. Only the financial model is refitted as its estimation window expands for OOS analysis.

\subsection{Transformer Text Layer: Frozen Sentence Transformer with
\texorpdfstring{$k$}{k}-Means}
\label{sec:methods:transformer}

The transformer branch replaces LDA's generative assignment with a
clustering of passage embeddings. We abbreviate it ST in the tables.
It produces the same two objects as Section~\ref{sec:methods:lda}, a
set of article weights $\theta_d$ on $L=60$ topics and a set of term
distributions used to describe those topics, so the stages that follow
consume it without modification. Throughout, $k$-means refers to the
clustering algorithm run with $k=L$ clusters.

The encoder is paraphrase-MiniLM-L3-v2, a Sentence-BERT model
\citep{reimers2019sentencebert} with three transformer layers, a
hidden size of 384, twelve attention heads, and a subword vocabulary
of 30,522 pieces. The representations of a passage's subword tokens
are averaged and the result is scaled to unit length. We use the
encoder only in the forward direction and never update its weights.
Freezing it serves two purposes. Training it on realized returns would
place the outcome we are trying to explain inside the text layer, which
would remove the separation between measuring narratives and evaluating
them. It would also make the comparison with LDA uninformative, since
LDA never sees returns. The question we are asking is whether sentence
representations learned without any financial supervision already
separate news narratives more sharply than word counts do.

Section~\ref{sec:methods:data} describes how articles are divided into
passages. The procedure yields 1,183,874 passages from 394,661
articles. Because the encoder accepts at most 128 subword tokens, a
passage longer than that is truncated, so each retained passage
contributes roughly its first 110 words of running text. The
consequences for the comparison are set out in
Section~\ref{sec:methods:fixed}.

We write $e_c \in \mathbb{R}^{384}$ for the unit-length embedding of
passage $c$ and fit $L$ cluster centers by mini-batch $k$-means with a
batch size of 4{,}096, ten initializations, and a random seed of zero.
Let $\mu_\ell$ denote the $\ell$th center scaled to unit length. Rather
than assigning each passage to its single closest center, we assign it
across centers in proportion to an exponential function of its cosine
similarity to each one,
\begin{equation}
    p_{c\ell}
    =\frac{\exp\!\bigl(e_c^{\top}\mu_\ell/\tau\bigr)}
          {\sum_{m=1}^{L}\exp\!\bigl(e_c^{\top}\mu_m/\tau\bigr)},
    \qquad \tau=0.07 ,
    \label{eq:soft-assign}
\end{equation}
where $\tau$ is a temperature that controls how concentrated the
assignment is. A small $\tau$ pushes almost all of the weight onto the
nearest center, while a large $\tau$ spreads it evenly. At $\tau=0.07$
the assignment is close to, but not exactly, a nearest-center rule, so
a passage that sits between two centers contributes to both.

Article weights follow by averaging over the article's passages. Let
$w_c$ be the word count of passage $c$ and let $C(d)$ be the set of
passages belonging to article $d$. Then
\begin{equation}
    \theta_{d\ell}
    =\frac{\sum_{c\in C(d)} w_c\,p_{c\ell}}
          {\sum_{c\in C(d)} w_c} ,
    \label{eq:doc-theta}
\end{equation}
which is nonnegative and sums to one across topics, the same support as
the LDA weights in Equation~\eqref{eq:lda}. Articles with no usable
passage receive equal weight $1/L$ on every topic. Daily attention is
then the term-count-weighted average of $\theta_d$ across the articles
published on that calendar day, the same aggregation applied to the LDA
branch in Section~\ref{sec:methods:lda}.

The clustering assigns articles to topics but does not describe them.
We therefore compute a class-based TF-IDF over the shared vocabulary
\citep{grootendorst2022bertopic}. Let $x_{dv}$ be the count of
vocabulary term $v$ in article $d$ and let $q_{d\ell}$ be the weight
with which article $d$ contributes its words to topic $\ell$, taken
here as one for the topic with the largest $\theta_{d\ell}$ and zero
otherwise. Writing $t_{\ell v}$ for the resulting term totals and
$\iota_v$ for a factor that discounts terms common to many topics,
\begin{equation}
    t_{\ell v} = \sum_{d} q_{d\ell}\,x_{dv} + 1,
    \qquad
    \iota_v = \log\!\left(1+
      \frac{L^{-1}\sum_{m,u}t_{mu}}{\sum_{m}t_{mv}}\right),
    \qquad
    \hat\phi_{v\ell} = \frac{t_{\ell v}\,\iota_v}
                            {\sum_{u} t_{\ell u}\,\iota_u} .
    \label{eq:ctfidf}
\end{equation}
Each topic is labeled by its three highest-weighted terms, and its ten
highest-weighted terms supply the NPMI calculation of
Section~\ref{sec:methods:metrics}. As with LDA, no topic names are
supplied to the model beforehand. We stress one difference from
Equation~\eqref{eq:lda}. The LDA term distribution $\phi_\ell$ is a
parameter estimated jointly with the article weights, whereas
$\hat\phi_\ell$ is computed after the clustering is complete and is
only a description of the vocabulary associated with each cluster. It
does not enter the construction of shocks, covariances, or factors. Only
$\theta_d$ propagates to the asset-pricing stages.

\subsection{What Is and Is Not Held Fixed}
\label{sec:methods:fixed}

The controlled comparison is defined by what the two branches share.
Both read the same 394,661 articles, both produce $L=60$ topics, both
deliver a daily attention series of the same shape to the same
construction of shocks, covariance characteristics, and factors, and
both are evaluated on the same 95,071 stock--month observations over the
same 41 evaluation months. The vocabulary of 15,000 terms is built once,
before either branch is fitted, so the coherence comparison in
Section~\ref{sec:methods:metrics} is conducted on a common set of
terms. A shared software interface checks that both branches return
weights of the correct shape that are nonnegative and sum to one before
any later stage runs.

The vocabulary plays different roles in the two branches, and this is
worth stating directly. For LDA the term counts are the model's input
and $\Phi$ is a parameter of the likelihood in
Equation~\eqref{eq:lda}. For the transformer the term counts enter only
through Equation~\eqref{eq:ctfidf}, after the clustering is finished,
and serve only to describe topics and to measure coherence. The two
branches are therefore matched on the basis used to evaluate term
lists, but not on the role that basis plays in estimation.

One input is not held fixed. LDA reads the full cleaned article,
whereas the transformer reads at most three opening passages, each
truncated by the encoder to roughly 110 words. The two branches
consequently see different amounts of each article, and any difference
we report reflects the change in representation together with this
change in the text supplied. We did not equalize the two, and we do not
attempt to separate their contributions. The rule used to rank terms
also differs, since LDA terms are ranked by the fitted $\phi_\ell$ and
transformer terms by Equation~\eqref{eq:ctfidf}, so the coherence
comparison in particular mixes the effect of the topic assignment with
the effect of the rule used to summarize it. Both limitations are
revisited in Section~\ref{sec:discussion}.

\subsection{Extension: Spherical Clustering}
\label{sec:methods:extension}

The specifications described here are not part of the controlled
comparison. They change the transformer text layer itself and are
reported separately in Table~\ref{tab:ladder}.

\paragraph{Spherical $k$-means.}
Passage embeddings are scaled to unit length, so what distinguishes two
passages is the angle between them rather than the distance. Mini-batch
$k$-means nevertheless minimizes squared Euclidean distance, and its
update sets each center to the ordinary mean of its members, which does
not lie on the unit sphere. Minimizing Euclidean distance and
maximizing cosine similarity coincide only when both vectors have unit
length, so once the centers drift off the sphere the two criteria
separate. We therefore replace the algorithm with spherical $k$-means
\citep{dhillon2001concept}, which can be read as the hard-assignment
counterpart of a mixture of von Mises--Fisher distributions sharing a
common concentration \citep{banerjee2005vmf}. Each passage is assigned
to the center with which it has the highest cosine similarity, and each
center is rescaled to unit length after every update,
\begin{equation}
    \max_{\{\mu_\ell\},\,\{a_c\}}\;
    \sum_{c} e_c^{\top}\mu_{a_c}
    \qquad\text{subject to}\qquad \|\mu_\ell\|_2=1 ,
    \label{eq:spherical-objective}
\end{equation}
where $a_c\in\{1,\dots,L\}$ is the cluster assigned to passage $c$.
Every passage is assigned before the centers are updated, so the block
size used to stream the embeddings affects memory use but not the
solution. A mini-batch $k$-means fit on a random sample of 100,000
passages supplies the starting centers. The run reported here converged
after 24 of a permitted 25 iterations, at a tolerance of $10^{-5}$ on
the average movement of the centers, reaching an average cosine
similarity of 0.5915 between passages and their assigned centers. Soft
assignment, article weights, and daily aggregation then follow
Equations~\eqref{eq:soft-assign} and~\eqref{eq:doc-theta} unchanged, at
the same temperature and the same seed.

\paragraph{Sparse-soft term head.}
Equation~\eqref{eq:ctfidf} accepts any nonnegative weight $q_{d\ell}$.
Giving an article entirely to its largest topic discards the rest of
its weight vector, while using $\theta_d$ itself lets every article
contribute its full word count to all 60 topics, which blurs the term
lists. The sparse-soft head sits between these. For each article let
$S_d$ hold the two topics with the largest weights and let
$m_d=\theta_{d(1)}-\theta_{d(2)}$ be the gap between the largest and
second-largest weights, which measures how decisively the article is
assigned. With $\kappa$ the median of $m_d$ across articles,
\begin{equation}
    q_{d\ell}
    =\mathbf{1}\{\ell\in S_d\}\;
     \frac{\theta_{d\ell}^{2}}{\sum_{m\in S_d}\theta_{dm}^{2}}\;
     m_d\,\mathbf{1}\{m_d\ge\kappa\} .
    \label{eq:sparse-soft}
\end{equation}
Squaring sharpens the contrast between the two retained topics, the
second indicator drops the half of the corpus whose assignment is least
decisive, and the factor $m_d$ weights the rest by how decisive they
are. Term counts additionally enter in the form $1+\log x_{dv}$ for
$x_{dv}>0$, so that a term repeated many times within a single article
does not dominate its topic's profile. In the reported run the median
gap is $\kappa=0.1013$, and the retained articles carry an average
total weight of 0.1423 across topics.

\paragraph{The two changes are separable.}
Equation~\eqref{eq:sparse-soft} governs only the description of topics.
Daily attention continues to use $\theta_d$ from
Equation~\eqref{eq:doc-theta} in full, so the input to the
asset-pricing stages is untouched. The two spherical specifications
were produced from a single set of cluster centers, identified by a
shared checkpoint fingerprint, and differ only in whether the term head
uses $\theta_d$ directly or Equation~\eqref{eq:sparse-soft}. Their
daily attention series agree across all 60 topics to a minimum
correlation of $1-6\times10^{-14}$, and their term distributions have
an average best-match cosine similarity of 0.887. The increase in mean
NPMI from 0.3702 to 0.4691 is therefore produced entirely by the
ranking of terms and cannot account for any part of the difference in
portfolio performance, a point developed in
Section~\ref{sec:methods:metrics}.

\subsection{Multi-Horizon IPCA}
\label{sec:methods:multihorizon}

The project's single-horizon baseline constructs one covariance
instrument per narrative and stock, using a 252-trading-day lookback
and a 60-trading-day exponential half-life. With $k=60$ narratives
and an intercept, the instrument vector $c_{i,t}$ has 61 entries.
Sparse IPCA maps these instruments into stock-specific loadings on
$K=3$ common factors \citep{kelly2019characteristics}, rather than
extracting principal components
of the attention series alone. The extension retains this
conditional factor structure but allows narrative exposures to
reflect different rates of change.

Here, \emph{multi-horizon} refers to the decay rates used to estimate
covariances, not to different return-forecast horizons. For each
$h\in\mathcal H=\{20,60,120\}$, we compute
$c^{(h)}_{i,t,\ell}=\operatorname{EWCov}_{h}(r^e_i,z_\ell)$,
where $z_\ell$ is the same trailing-five-observation attention shock
at every horizon and $r^e_i$ is the daily stock return minus the
daily risk-free return. All three estimates use the same
252-trading-day lookback ending before the priced month, require at
least 120 usable observations, and assign an observation $s$
trading days before the window end a weight proportional to
$2^{-s/h}$. Thus, the short half-life emphasizes recent comovement,
whereas the long half-life retains more of the historical exposure.
The extension changes the weighting of the observations, not the
length of the window.

We concatenate the three sets of 60 instruments and include an
intercept. After column scaling, $c_{i,t}$ therefore has 181
entries, and the loading map $\Gamma$ is $181\times3$. The model is
\begin{equation}
    r^e_{i,t+1}=c_{i,t}\Gamma f_{t+1}+e_{i,t+1}.
    \label{eq:multihorizon-ipca}
\end{equation}
The return horizon remains one month, but the extension uses excess
rather than total returns: the compounded monthly total stock return
minus the compounded monthly risk-free return. Monthly stock returns
require at least 15 daily observations. Each horizon has its own
coefficients in $\Gamma$, but all
horizons share the same three-dimensional factor return $f_{t+1}$.
No equality restriction is imposed on the horizon-specific
coefficients. The 180 covariance instruments are consequently
neither 180 latent factors nor inputs to three separate models.

In the baseline, each narrative contributes one penalized row of
$\Gamma$. The extension instead imposes sparsity jointly across
that narrative's horizons. Let $\Gamma_\ell$
collect the three horizon rows associated with narrative $\ell$,
forming a $3\times3$ block. The implementation minimizes
\begin{equation}
    \min_{\Gamma,\{f_{t+1}\}}
    \sum_{i,t}\bigl(r^e_{i,t+1}-c_{i,t}\Gamma f_{t+1}\bigr)^2
    +\lambda N\sum_{\ell=1}^{60}\|\Gamma_\ell\|_F
    +\sum_t\|f_{t+1}\|_2^2,
    \label{eq:multihorizon-objective}
\end{equation}
where $N$ is the number of training stock--month observations.
The intercept row is unpenalized. Alternating updates estimate
factor returns by ridge regression conditional on the loadings
and update $\Gamma$ with a grouped penalty conditional on the
factors. A zero block removes a narrative at all three horizons;
a retained block can assign different weights to its horizons.
This structure allows recent and persistent exposures to enter
jointly without increasing $K$, while preserving a
narrative-level selection rule.

\subsection{Evaluation and Results}
\label{sec:methods:metrics}

\paragraph{Sample and estimation.}
The FNSPID comparison uses 394,661 articles dated 2015--2023 and a
financial panel of 95,071 stock--month observations spanning 102
months, August 2015--January 2024. All specifications in
Table~\ref{tab:ladder} use $k=60$ narratives and $K=3$ factors.
The first 61 months form the initial estimation sample; the
remaining 41 months, September 2020--January 2024, supply the
reported out-of-sample (OOS) portfolio returns. The model is
re-estimated on an expanding sample every 12 months, starting in
September 2020. Within each estimation sample, the last 25\% of
months are reserved for chronological validation of $\lambda$.
Selection maximizes validation portfolio Sharpe, preferring the
largest penalty among scores within a tolerance of 5\% of the
absolute best score.
The selected penalty is then used to refit on the entire
estimation sample. The single-horizon runs search 12 penalty
values and the multi-horizon runs search eight.

\paragraph{Metrics and portfolio construction.}
Topic coherence is measured using document-level co-occurrence
in the shared vocabulary. For terms $a,b$, normalized pointwise
mutual information is
\begin{equation}
    \operatorname{NPMI}(a,b)
    =\frac{\log[p(a,b)/(p(a)p(b))]}{-\log p(a,b)}.
    \label{eq:npmi}
\end{equation}
Here $p(a)$ and $p(a,b)$ are the proportions of documents containing
term $a$ and both terms, respectively.
The implementation uses a small numerical floor for zero
probabilities. We average over the 45 pairs among each topic's
top ten terms and then over the 60 topics. Higher NPMI indicates
more coherent term lists under this criterion; it is not a
measure of return predictability or a validation of the topics'
economic interpretation.

Portfolio weights in factor space are computed from the training
factor returns as $w=\Sigma_f^{-1}\mu_f$, where $\mu_f$ and
$\Sigma_f$ are their estimated mean and covariance. For an OOS
month, let $B_{t+1}$ contain the stock loadings $c_{i,t}\Gamma$,
constructed from lagged covariance instruments. The realized factor return
and portfolio return
are
\begin{equation}
    \widehat f_{t+1}
    =(B_{t+1}'B_{t+1}+I_K)^{-1}B_{t+1}'r_{t+1},
    \qquad R_{t+1}=w'\widehat f_{t+1}.
    \label{eq:oos-portfolio}
\end{equation}
Here $r_{t+1}$ is the vector of total returns in the original runs
and excess returns in the extensions. Realized stock returns
are used to measure the month's factor return, not to choose
$w$ retrospectively in the financial evaluation step.
Equivalently, conditional on the fitted parameters and instrument scales,
the stock portfolio has weights
$B_{t+1}(B_{t+1}'B_{t+1}+I_K)^{-1}w$. The reported annualized ratio is
$\overline R/\operatorname{sd}(R)\times\sqrt{12}$, using the sample
standard deviation of the 41 monthly observations.

\begin{table}[htbp]
\centering
\small
\caption{Development sequence on FNSPID news, $k=60$ and $K=3$.
OOS Sharpe ratios use September 2020--January 2024 (41 months).
Half-lives are in trading days. Rows are successive specifications,
not a component-by-component ablation.}
\label{tab:ladder}
\begin{tabular}{@{}p{0.40\linewidth}lccc@{}}
\toprule
Specification & Returns & Half-lives & Mean NPMI & OOS Sharpe \\
\midrule
LDA, single-horizon IPCA
    & Total & 60 & 0.2875 & $-0.0810$ \\
Frozen ST, single-horizon IPCA
    & Total & 60 & 0.3469 & 0.3059 \\
Frozen ST, multi-horizon IPCA
    & Excess & 20/60/120 & 0.3469 & 0.6605 \\
Spherical ST, sparse-soft c-TF-IDF, multi-horizon IPCA
    & Excess & 20/60/120 & 0.4691 & 1.0334 \\
\bottomrule
\end{tabular}
\end{table}

\paragraph{Empirical patterns.}
Replacing LDA with the original frozen ST text layer raises mean
NPMI by 0.0594 and the OOS ratio by 0.3869. This comparison keeps
the downstream financial specification fixed, although LDA
uses full articles whereas ST uses lead chunks
(Section~\ref{sec:methods:fixed}). It supports a descriptive
improvement in term coherence and portfolio performance in this
sample, not an isolated estimate of the effect of contextual
encoding. The reported equal-weight market portfolio has an OOS
total-return ratio of 0.8166 on the same window, above both
single-horizon narrative portfolios; this is not a like-for-like
benchmark for the excess-return extensions. The separate
replication using the authors' supplied
LDA attention and historical sample is not a same-window
benchmark for this FNSPID comparison.

The move from single-horizon ST to multi-horizon ST raises the
reported OOS ratio from 0.3059 to 0.6605, a difference of 0.3546.
NPMI is unchanged because the text layer is unchanged. This step
also switches from total to excess returns and changes the
penalty search grid, so the table cannot attribute the entire
difference to the covariance half-lives. The three half-lives
enter jointly; the reported results do not establish which
half-life contributes most or whether each improves performance
on its own.

The spherical specification raises the OOS excess-return Sharpe
by a further 0.3729, to 1.0334, relative to the preceding
multi-horizon specification. Its mean NPMI rises by 0.1222.
These two improvements concern different parts of the pipeline.
For the spherical attention assignments, the full-soft term
head yields NPMI 0.3702, compared with 0.4691 for the sparse-soft
head. The difference of 0.0989 therefore reflects term ranking
with effectively unchanged pricing attention, rather than a
change in the return signal. The sparse-soft head therefore cannot
explain the Sharpe gain. Spherical clustering changes the
attention representation, but the accompanying increase in
coherence is not evidence that coherence itself causes higher
portfolio performance.

The grouped penalty also does not produce a persistently sparse
multi-horizon model in these runs. Original multi-horizon ST
retains 34 narratives at the first OOS refit and all 60 at the
subsequent refits; spherical ST retains all 60 at each refit.
Thus, the higher Sharpe is not accompanied by a smaller selected
narrative set. Group norms summarize the importance of
narratives in the fitted loading map, not unique names for the
three latent factors, whose individual interpretation depends
on the chosen factor rotation.

\paragraph{Uncertainty and evaluation boundaries.}
The available tests do not establish a Sharpe improvement.
For the original ST--LDA comparison, a paired test of mean OOS
portfolio returns gives $p=0.22$. The topic-level Welch test
gives $p=0.042$, but topics extracted from the same corpus need
not be independent; this is not a clean test of the text
architecture. For spherical ST versus the preceding
multi-horizon ST run, the paired mean-return test gives
$t=0.692$ and $p=0.493$; a Newey--West mean-difference test
with three monthly lags gives $p=0.521$. A reported paired block bootstrap
(20,000 replications, six-month blocks) yields a 95\% interval
of $[-0.903,1.682]$ for the Sharpe difference. This interval
includes zero and is wide relative to the point difference.
These mean-return tests and Sharpe intervals answer different
questions, and neither supports a claim of statistically
established outperformance.

The raw spherical portfolio has annualized volatility of
81.68\% and a maximum drawdown of 62.23\%. The presentation's
10\%-volatility plots rescale returns for display using an
ex-post scale; they are not the returns of a prospectively
risk-managed, cost-adjusted strategy. The observed Sharpe gain
therefore does not establish superior implementable performance
under a common leverage or trading-cost budget.

Finally, the OOS designation applies to the financial fitting
schedule, not to every stage of the pipeline. Instrument
columns are divided by standard deviations computed over the
full panel, including OOS months; the vocabulary and clustering
representations are built using the full text sample; and the
extension was selected after inspecting the same 41 OOS
months. These choices admit future-sample information or
selection effects despite correctly lagged shocks, covariance
windows, and financial refits. The results are exploratory
comparisons conditional on these choices, rather than estimates
from a fully prospective holdout.

% ---------------------------------------------------------------------
\section{Results}
\label{sec:results}

The project investigates whether a frozen contextual text
representation can replace LDA in a narrative asset-pricing
pipeline without changing its conditional factor-model logic.
The original comparison produces more coherent term lists and
a higher OOS portfolio ratio for ST on the same financial
window. The extensions then combine a revised attention
representation with multiple covariance decay rates. The
largest observed excess-return Sharpe belongs to spherical ST
with grouped multi-horizon IPCA, but the uncertainty reported in
Section~\ref{sec:methods:metrics} prevents treating this ranking
as an established performance advantage.

The useful methodological distinction is between narrative
measurement and stock exposure. Spherical clustering changes
how articles contribute to daily attention, while the
multi-horizon instruments let each stock's loadings depend on
both recent and more persistent comovement with the same
narrative shocks. The sparse-soft term head improves the
coherence of the displayed labels without changing those
pricing assignments. Consequently, improved labels and
improved portfolio performance are separate findings.
Moreover, retaining all narratives at the later refits limits
the claim that group sparsity identifies a small set of priced
topics. The specification offers an interpretable map from
narrative exposures to factor loadings, not three uniquely
identified economic narratives.

The evidence remains limited by a short, 41-month evaluation
window, specification search on that window, and full-sample
text construction and instrument scaling. Unequal article
spans in the LDA and ST arms and different term-ranking rules
further limit attribution of the coherence differences. The
reported runs fix $k=60$ and the clustering seed rather than
measuring sensitivity to either choice. Missing intraday
article timestamps also prevent verification that each day's
news would have been available before the trading decisions
that a live strategy would require.

An informative follow-up would freeze the text representation,
training-only instrument scales, and specification choices
before an untouched evaluation period. Equalizing the text
spans and term-ranking rule would sharpen the comparison of
text architectures; matched runs changing only the covariance
half-lives would isolate the multi-horizon contribution.
Repeating those comparisons across clustering seeds and topic
counts would assess whether the ranking depends on a single
partition. Within the present design, the findings motivate
these tests, but do not establish that the extensions outperform
the original paper or deliver a cost-adjusted trading advantage.

% ---------------------------------------------------------------------
\section{Discussion}
\label{sec:discussion}

\paragraph{Limitations.}

The LDA and transformer topic models both use the same collection of articles to produce 60 topics. However, the LDA receives the complete article text, post-cleaning, whereas the transformer model receives up to three opening chunks. Therefore, it is not an exact ``apples-to-apples'' comparison in the sense that differences in performance may be driven by both the modeling method and the text available to each model. Furthermore, the topic count is fixed throughout our analysis, leaving its influence on performance and results unresolved.

NPMI measures how often a topic's selected top words appear together across articles. In our analysis, changing which words appear in that list can improve NPMI without actually changing how articles are assigned to each topic. As a result, changes in the words used to describe each topic can improve NPMI without changing the underlying attention series. Thus, higher coherence does not necessarily indicate better asset-pricing signals. Our NPMI scores cannot be directly compared with studies using different measures for coherence, such as $c_v$ \citep{roder2015coherence}.

Additionally, the financial evaluation is limited to 41 months, and the OOS tests performed do not establish a statistically significant improvement in Sharpe ratios.\footnote{See Section~\ref{sec:methods:metrics} for the reported tests and confidence interval.}

\paragraph{Future work.}

A closer comparison would give both models the same article passages and use a common procedure for ranking topic terms. Varying the number of topics and repeating estimations with several random seeds would help establish whether the findings depend on a particular topic partition. Permutation tests that preserve temporal dependence could also provide further evidence on whether the relation between news attention and returns is stronger than would be expected by chance.

A further step is to evaluate the complete procedure using only information available up to each estimation date. Currently, both the LDA and transformer models use the full news corpus from January 2015 to December 2023 to construct topics. The vocabulary and topic models could be estimated before the evaluation period and then held fixed as new articles arrive, with covariance scaling based only on the financial training sample. Model choices should be settled before testing on an untouched later period. Evaluating all portfolios on an excess-return basis would also make their performance comparable. Finally, GRS tests \citep{gibbons1989test} on anomaly portfolios could assess whether the estimated factors jointly explain average returns, extending the evaluation beyond the performance of the factors' own portfolios.\footnote{The GRS test evaluates the joint null hypothesis that all test-portfolio intercepts are zero in regressions of excess returns on the factors. Failure to reject this null does not establish that the model prices all assets correctly.}

% ---------------------------------------------------------------------
\section{Conclusion}
\label{sec:conclusion}

The construction of narrative factors in our analysis comes down to how well financial news is being summarized. We examine the construction of narrative factor models, using financial news, through the asset-pricing framework laid out in Bybee et~al.\ (2023) \citep{bybee2023narrative}, replacing the LDA topic model with a frozen sentence transformer. Both the LDA and transformer models use the same collection of FNSPID articles, and both produce 60 topics in total. With the topic count and shared vocabulary held fixed across models, the transformer model yields higher topic coherence and OOS Sharpe ratios. Our comparison leaves some differences unresolved, most notably the short article passages being used by the transformer.

Our extensions of Bybee et~al.\ (2023) examine spherical clustering and stock exposures that place different weights on older and newer news shocks. This combined specification attained a Sharpe ratio of approximately 1.03, which was the highest among the model specifications considered. Although the results outlined in the paper are not sufficient to clearly establish outperformance, given the 41-month evaluation period, these estimates do provide a compelling reason to investigate the approach further.

The economic interpretation of the transformer model continues to rest on news narratives. With our model, we still examine the terms associated with a topic, measure changes in its attention, and relate those changes to movements in stock returns. This preserves the central theme of \textit{Narrative Asset Pricing: Interpretable Systematic Risk Factors from News Text} \citep{bybee2023narrative}, even as the topic model used becomes more complex. The evidence presented in this paper favors further study of contextual financial news representations as inputs for an interpretable asset-pricing model \citep{bell2026glassbox,bell2026alphaglass}.

% ---------------------------------------------------------------------
\section{Data and Code Availability}
\label{sec:availability}

The code, configuration files, and shareable artifacts used in this study are available in the \href{https://github.com/Shivesh777/bert-lda}{project's GitHub repository}. The repository includes instructions for rebuilding the corpus, fitting each text model, constructing the financial characteristics, and reproducing the reported tables and evaluation summaries. It also contains the derived topic labels, attention weights, model outputs, and other results that can be redistributed.

The financial-news sample is drawn from the public FNSPID dataset \citep{dong2024fnspid}, distributed through \href{https://huggingface.co/datasets/Zihan1004/FNSPID}{Hugging Face}. We apply the filtering, deduplication, and preprocessing rules described in Section~\ref{sec:methods:data}. Daily stock returns and market-capitalization data come from the CRSP US Stock Databases, accessed through the Wharton Research Data Services platform \citep{crspstockdatabase}. CRSP data are licensed and therefore are not redistributed; the repository provides a script that reconstructs the study universe for users with the required WRDS access. The Fama--French daily factor file used to obtain the risk-free rate is publicly available from the \href{https://mba.tuck.dartmouth.edu/pages/faculty/ken.french/data_library.html}{Kenneth R. French Data Library}.

\clearpage
% ---------------------------------------------------------------------
% Bibliography embedded so main.tex compiles without a separate .bib file
% or BibTeX pass. The editable BibTeX database remains at the top of this
% source for reference maintenance.

% ---------------------------------------------------------------------
\clearpage
\appendix
\numberwithin{table}{section}
\section{Additional Tables and Figures}
\label{app:extra}

\begin{table}[!htbp]
\centering
\renewcommand{\tablename}{Exhibit}
\caption{Preprocessing for the LDA and transformer models.}
\label{tab:preprocessing}
\small
\setlength{\tabcolsep}{4pt}
\renewcommand{\arraystretch}{1.18}
\begin{tabular}{@{}p{0.17\textwidth}p{0.38\textwidth}p{0.38\textwidth}@{}}
\toprule
\textbf{Step} & \textbf{LDA} & \textbf{Transformer} \\
\midrule
Starting text & Headline combined with the article body, capped at 4,000 characters before combination. & Same retained headline and article body. \\
\addlinespace
Text cleaning & Remove URLs and nonalphabetic characters, convert to lowercase, and normalize spacing. & Preserve the wording, capitalization, and punctuation before dividing the text into chunks. \\
\addlinespace
\begin{tabular}[t]{@{}l@{}}Vocabulary\\selection\end{tabular} & Select single words and word pairs from 60,000 randomly sampled articles, using seed 0. Exclude English stopwords and financial or publisher terms. Candidate words have at least three letters. Terms must occur in at least 10 sampled articles and no more than 40\% of the sample. & Use the same vocabulary for topic descriptions and coherence evaluation. It does not restrict the encoder's tokenizer or the text supplied to it. \\
\addlinespace
Text used & Count retained vocabulary terms across the complete cleaned headline and capped article body. & Split at paragraph boundaries and divide longer paragraphs into chunks of at most 200 words. Exclude chunks shorter than eight words and retain up to three leading eligible chunks per article. \\
\addlinespace
Model input & A sparse article--term count matrix with 15,000 columns, comprising 7,467 single words and 7,533 word pairs. & Encode each retained chunk with the frozen paraphrase-MiniLM-L3-v2 sentence transformer. Normalize each embedding to unit length. Input remains subject to the encoder's token limit. \\
\bottomrule
\end{tabular}
\vspace{0.6em}
\begin{minipage}{0.96\textwidth}
\footnotesize
\textit{Notes:} Both branches retain a common article index and publication dates. Stopwords are excluded during vocabulary selection. The subsequent count pass uses the fixed vocabulary without a separate stopword-removal step. Vocabulary selection draws from the full news period, rather than only the initial financial estimation sample. The shorter text supplied to the transformer remains a difference between the two models.
\end{minipage}
\end{table}

%\todo{Full topic lists for both arms.}

% ---------------------------------------------------------------------
\clearpage
\section{Ablation Study: Number of Topics}
\label{app:ablation}

The main text fixes $L=60$ topics for both branches. To check the
sensitivity of the results to that choice, we refit both text layers
at $L\in\{50,60,70,80\}$ and compare coherence at every $L$, and we
carry the $L=80$ fits through the full pricing pipeline. Everything
downstream of the text layer is unchanged: the same encoder, seed, and
Gibbs schedule, the same vocabulary construction rule, the same
95,071 stock--month panel, $K=3$ factors, the same penalty grids, and
the same 41 evaluation months.

\begin{table}[!ht]
\centering
\small
\caption{Coherence of the two text layers as the number of topics
varies. NPMI follows Equation~\eqref{eq:npmi} over the shared
vocabulary and the ten highest-ranked terms per topic. Diversity is
the share of distinct terms among the $L\times 10$ top terms.}
\label{tab:ablation-coherence}
\vspace{0.6em}
\setlength{\tabcolsep}{6pt}
\begin{tabular}{@{}llrr@{}}
\toprule
Text layer & $L$ & Mean NPMI & Diversity \\
\midrule
LDA                              & 50 & 0.2867 & 0.736 \\
                                 & 60 & 0.2875 & 0.683 \\
                                 & 70 & 0.2910 & 0.639 \\
                                 & 80 & 0.2824 & 0.606 \\
\addlinespace
Spherical ST, soft head          & 50 & 0.3440 & 0.448 \\
                                 & 60 & 0.3702 & 0.480 \\
                                 & 70 & 0.3791 & 0.449 \\
                                 & 80 & 0.4009 & 0.470 \\
\addlinespace
Spherical ST, sparse-soft head   & 50 & 0.4414 & 0.700 \\
                                 & 60 & 0.4691 & 0.697 \\
                                 & 70 & 0.4622 & 0.669 \\
                                 & 80 & 0.4943 & 0.666 \\
\bottomrule
\end{tabular}
\vspace{-1.2em}
\end{table}

\paragraph{Coherence.}
The ordering of the two text layers does not depend on $L$: the
sparse-soft spherical ST head exceeds LDA on mean NPMI by 0.155 to
0.212 at every topic count. Within each layer, $L$ matters little.
LDA moves between 0.282 and
0.291 with no monotone pattern, and spherical ST drifts from 0.441 to
0.494, a difference that is not statistically significant and that
comes with lower diversity, so the extra topics partly reuse the
terms of existing ones.

\newpage
\begin{table}[!ht]
\centering
\small
\caption{Pricing results at $L=80$ and the corresponding $L=60$ runs
of Table~\ref{tab:ladder}. OOS Sharpe ratios use September 2020 to
January 2024 (41 months) and $K=3$. Half-lives are in trading days.
Active is the number of narratives retained at the final refit. The
single-horizon $L=60$ transformer row uses the original frozen ST
clustering, whereas all $L=80$ transformer rows use spherical ST with the
sparse-soft head.}
\label{tab:ablation-pricing}
\vspace{0.6em}
\setlength{\tabcolsep}{5pt}
\begin{tabular}{@{}p{0.34\linewidth}llcccc@{}}
\toprule
Specification & Returns & Half-lives & \multicolumn{2}{c}{OOS Sharpe} & \multicolumn{2}{c}{Active} \\
\cmidrule(lr){4-5}\cmidrule(lr){6-7}
 & & & $L=60$ & $L=80$ & $L=60$ & $L=80$ \\
\midrule
Spherical ST, sparse-soft, multi-horizon IPCA
    & Excess & 20/60/120 & 1.0334 & 0.7156 & 60 & 80 \\
LDA, multi-horizon IPCA
    & Excess & 20/60/120 & --- & 0.6125 & --- & 80 \\
\addlinespace
ST, single-horizon IPCA
    & Total & 60 & 0.3059 & 0.7219 & 53 & 80 \\
LDA, single-horizon IPCA
    & Total & 60 & $-0.0810$ & 0.7675 & 60 & 78 \\
\addlinespace
Spherical ST, sparse-soft, single-horizon IPCA
    & Excess & 60 & --- & 0.6412 & --- & 80 \\
LDA, single-horizon IPCA
    & Excess & 60 & --- & 0.6470 & --- & 78 \\
\midrule
Equal-weight market, same universe
    & Total & & \multicolumn{2}{c}{0.8166} & & \\
Equal-weight market, same universe
    & Excess & & \multicolumn{2}{c}{0.7164} & & \\
\bottomrule
\end{tabular}
\vspace{-1.2em}
\end{table}

\paragraph{Pricing.}
At $L=80$ the headline specification delivers an OOS excess-return
Sharpe ratio of 0.7156, against 1.0334 at $L=60$, and still exceeds
its LDA counterpart (0.6125). With 41 monthly observations the
standard error of an annualized Sharpe ratio is roughly
$\sqrt{12/41}\approx 0.55$, so neither the decline from $L=60$ to
$L=80$ nor the gap between the text layers at $L=80$ is
statistically distinguishable from zero, consistent with the
bootstrap interval in Section~\ref{sec:methods:metrics}. On the
single-horizon specification the two text layers are
indistinguishable at $L=80$ and both sit slightly below the
equal-weight market. More topics do not help the pricing stage
because the grouped penalty selects almost nothing at $L=80$
(penalties of order $10^{-5}$ to $10^{-7}$, 77 to 80 of 80
narratives active) while the multi-horizon panel grows from 180 to
240 instruments on the same observations. The in-sample Sharpe ratio
of the spherical ST fit falls from 1.34 at the first refit to 0.87 at
the last, the usual signature of a larger instrument set fitted to a
fixed panel.

\paragraph{Conclusion.}
The sweep gives no reason to move away from $L=60$. The coherence
advantage of the transformer branch holds at every topic count we
tried, so the choice of $L=60$ does not favor either branch, and
raising $L$ to 80 lowers diversity, removes what little sparsity the
grouped penalty produced, and lowers the point estimate of the
headline Sharpe ratio without changing any conclusion in the main
text. We therefore retain $L=60$ and treat the $L=80$ results as a
robustness check.

% \section{Implementation Details}
% \label{app:impl}
% \todo{Hyperparameters, software versions, compute.}

\end{document}

%% file: figures/fig1_pipeline.tex
\begin{tikzpicture}[
    >=Stealth,
    font=\sffamily,
    clipping/.style={draw=bordergray!85, fill=white, rounded corners=0.5pt, line width=0.4pt, font=\fontsize{5}{6}\selectfont},
    panel/.style={draw=bordergray!70, fill=white, rounded corners=3.5pt, line width=0.6pt},
    header/.style={font=\bfseries\footnotesize\color{navyblue}},
    annot/.style={font=\fontsize{6.5}{7.5}\selectfont\color{midgray}},
    dataflow/.style={->, line width=1.1pt, color=navyblue}
]

% =========================================================================
% TOP PANEL: STAGE 1 & 2 (News -> Embedding -> Attention & Shocks)
% Dimensions: x from -0.1 to 16.1, y from 5.80 to 12.20 (height = 6.40)
% =========================================================================
\draw[panel] (-0.1, 5.80) rectangle (16.1, 12.20);
\node[anchor=north west, font=\bfseries\scriptsize\color{navyblue}] 
    at (0.1, 12.02) {\textsc{Stage 1 \& 2: Financial News Corpus $\longrightarrow$ Semantic Narrative Attention \& Shocks}};

% -------------------------------------------------------------------------
% 1. FINANCIAL NEWS STREAM (x: 0.1 to 4.3)
% -------------------------------------------------------------------------
\node[header, anchor=north west] at (0.2, 11.55) {1. Financial News Stream};
\node[annot, anchor=north west] at (0.2, 11.20) {394k articles (FNSPID 2015--2023)};

% Newspaper Clipping 1: Fed / Rates
\node[clipping, text width=4.1cm, inner sep=2.5pt, anchor=north west] (card1) at (0.2, 10.75) {
    {\fontsize{4.2}{5.0}\selectfont\textsf{\textbf{\color{navyblue}MACRO / RATES}}\hfill\color{midgray}\textsf{FEB 2015}}\\[1pt]
    {\fontsize{5.2}{6.2}\selectfont\rmfamily\textbf{Fed Signals Gradual Liftoff}}\\[1pt]
    {\color{bordergray!50}\rule{\linewidth}{0.25pt}}\\[1.5pt]
    {\color{bordergray!40}\rule{2.6cm}{0.8pt}}\hspace{2pt}{\color{bordergray!25}\rule{0.9cm}{0.8pt}}\\[1.5pt]
    {\color{bordergray!40}\rule{1.9cm}{0.8pt}}\hspace{2pt}{\color{bordergray!25}\rule{1.5cm}{0.8pt}}
};

% Newspaper Clipping 2: Energy
\node[clipping, text width=4.1cm, inner sep=2.5pt, below=7pt of card1] (card2) {
    {\fontsize{4.2}{5.0}\selectfont\textsf{\textbf{\color{accentorange}ENERGY / COMMODITIES}}\hfill\color{midgray}\textsf{JUN 2016}}\\[1pt]
    {\fontsize{5.2}{6.2}\selectfont\rmfamily\textbf{Crude Falls on Supply Surplus}}\\[1pt]
    {\color{bordergray!50}\rule{\linewidth}{0.25pt}}\\[1.5pt]
    {\color{bordergray!40}\rule{2.4cm}{0.8pt}}\hspace{2pt}{\color{bordergray!25}\rule{1.1cm}{0.8pt}}\\[1.5pt]
    {\color{bordergray!40}\rule{2.0cm}{0.8pt}}\hspace{2pt}{\color{bordergray!25}\rule{1.4cm}{0.8pt}}
};

% Newspaper Clipping 3: Tech
\node[clipping, text width=4.1cm, inner sep=2.5pt, below=7pt of card2] (card3) {
    {\fontsize{4.2}{5.0}\selectfont\textsf{\textbf{\color{accentteal}TECHNOLOGY}}\hfill\color{midgray}\textsf{OCT 2015}}\\[1pt]
    {\fontsize{5.2}{6.2}\selectfont\rmfamily\textbf{Chipmakers Face Strong Demand}}\\[1pt]
    {\color{bordergray!50}\rule{\linewidth}{0.25pt}}\\[1.5pt]
    {\color{bordergray!40}\rule{2.7cm}{0.8pt}}\hspace{2pt}{\color{bordergray!25}\rule{0.8cm}{0.8pt}}\\[1.5pt]
    {\color{bordergray!40}\rule{1.7cm}{0.8pt}}\hspace{2pt}{\color{bordergray!25}\rule{1.6cm}{0.8pt}}
};

% Arrow 1 -> 2
\draw[dataflow] (4.45, 9.0) -- node[above, font=\fontsize{5.5}{6.5}\selectfont\bfseries, color=navyblue] {Passages} (5.80, 9.0);

% -------------------------------------------------------------------------
% 2. SEMANTIC REPRESENTATION (x: 5.3 to 9.8)
% -------------------------------------------------------------------------
\node[header, anchor=north west] at (5.5, 11.55) {2. Semantic Representation};
\node[annot, anchor=north west] at (5.5, 11.20) {Frozen Sentence-BERT embeddings};

% Projected 3D Unit Sphere
\begin{scope}[shift={(7.1, 9.0)}]
    % Outer boundary of unit sphere
    \draw[line width=0.8pt, color=navyblue!65] (0,0) circle (1.10);
    
    % Equatorial ellipse (3D depth cue)
    \draw[dashed, thin, color=bordergray!90] (-1.10, 0) arc (180:0:1.10 and 0.30);
    \draw[line width=0.45pt, color=bordergray] (-1.10, 0) arc (180:360:1.10 and 0.30);
    
    % Longitudinal meridian ellipse
    \draw[dashed, thin, color=bordergray!60] (0, 1.10) arc (90:270:0.30 and 1.10);
    \draw[line width=0.4pt, color=bordergray!80] (0, 1.10) arc (90:-90:0.30 and 1.10);

    % Unit norm annotation
    \node[font=\fontsize{5}{6}\selectfont, color=midgray] at (0.68, -0.72) {$\|\mathbf{e}\|_2 = 1$};

    % Topic cluster 1: Fed/Macro (blue)
    \fill[navyblue] (0.24, 1.00) circle (1.5pt);
    \fill[navyblue] (0.38, 0.94) circle (1.5pt);
    \fill[navyblue] (0.10, 1.02) circle (1.5pt);
    \draw[->, line width=1.1pt, color=navyblue] (0,0) -- (0.26, 0.98) 
        node[above right, font=\fontsize{5.5}{6.5}\selectfont\bfseries\color{navyblue}] {$\boldsymbol{\mu}_{\text{Fed}}$};

    % Topic cluster 2: Oil/Energy (orange)
    \fill[accentorange] (0.95, -0.28) circle (1.5pt);
    \fill[accentorange] (0.90, -0.40) circle (1.5pt);
    \fill[accentorange] (1.00, -0.15) circle (1.5pt);
    \draw[->, line width=1.1pt, color=accentorange] (0,0) -- (0.95, -0.26) 
        node[below right, font=\fontsize{5.5}{6.5}\selectfont\bfseries\color{accentorange}] {$\boldsymbol{\mu}_{\text{Oil}}$};

    % Topic cluster 3: Tech/Hardware (teal)
    \fill[accentteal] (-0.85, 0.36) circle (1.5pt);
    \fill[accentteal] (-0.75, 0.50) circle (1.5pt);
    \fill[accentteal] (-0.95, 0.16) circle (1.5pt);
    \draw[->, line width=1.1pt, color=accentteal] (0,0) -- (-0.85, 0.36) 
        node[above left, font=\fontsize{5.5}{6.5}\selectfont\bfseries\color{accentteal}] {$\boldsymbol{\mu}_{\text{Tech}}$};
\end{scope}

% Unboxed geometric annotation beneath sphere
\node[font=\fontsize{5.5}{6.5}\selectfont, color=midgray, align=center] at (7.1, 7.05) {
    Directional clustering on $\mathbb{S}^{d-1}$ ($L=60$)\\[1pt]
    \textit{\fontsize{4.8}{5.8}\selectfont Baseline: LDA word counts}
};

% Arrow 2 -> 3
\draw[dataflow] (8.35, 9.0) -- node[above, font=\fontsize{5.5}{6.5}\selectfont\bfseries, color=navyblue] {Attention} 
                                node[below, font=\fontsize{5}{6}\selectfont, color=midgray] {$\boldsymbol{\theta}_t \in \Delta^{60}$} (9.95, 9.0);

% -------------------------------------------------------------------------
% 3. DAILY ATTENTION & SHOCKS (x: 10.3 to 15.9)
% -------------------------------------------------------------------------
\node[header, anchor=north west] at (10.4, 11.55) {3. Daily Attention \& Shocks};
\node[annot, anchor=north west] at (10.4, 11.20) {Trailing shock filter: $\mathbf{z}_t = \boldsymbol{\theta}_t - \overline{\boldsymbol{\theta}}_{t-5:t-1}$};

% Attention time-series plot
\begin{scope}[shift={(10.5, 8.95)}]
    \draw[->, thin, color=midgray] (0,0) -- (5.1, 0) node[right, font=\fontsize{6}{7}\selectfont] {$t$};
    \draw[->, thin, color=midgray] (0,0) -- (0, 1.20) node[above, font=\fontsize{6}{7}\selectfont] {$\theta_{t,\ell}$};
    
    % Attention curve Fed (navy)
    \draw[line width=1.1pt, color=navyblue] plot[smooth, tension=0.7] coordinates {
        (0.2, 0.20) (0.9, 0.28) (1.6, 0.98) (2.3, 0.46) (3.1, 0.38) (4.0, 0.28) (4.8, 0.42)
    };
    \node[font=\fontsize{5.5}{6.5}\selectfont\bfseries, color=navyblue] at (1.6, 1.10) {Fed/Rates};

    % Attention curve Oil (orange)
    \draw[line width=1.1pt, color=accentorange] plot[smooth, tension=0.7] coordinates {
        (0.2, 0.46) (0.9, 0.38) (1.6, 0.20) (2.5, 0.24) (3.3, 0.92) (4.0, 0.54) (4.8, 0.20)
    };
    \node[font=\fontsize{5.5}{6.5}\selectfont\bfseries, color=accentorange] at (3.3, 1.04) {Oil/Energy};

    % 5-day rolling mean illustration (dashed gray)
    \draw[line width=0.8pt, dashed, color=midgray] plot[smooth, tension=0.7] coordinates {
        (0.4, 0.24) (1.0, 0.26) (1.6, 0.48) (2.3, 0.56) (3.1, 0.42) (4.0, 0.32) (4.8, 0.35)
    };
    \node[font=\fontsize{5}{6}\selectfont, color=midgray] at (4.3, 0.15) {trailing 5d MA};
\end{scope}

% Shock impulses plot below
\begin{scope}[shift={(10.5, 7.40)}]
    \draw[->, thin, color=midgray] (0,0) -- (5.1, 0) node[right, font=\fontsize{6}{7}\selectfont] {$t$};
    \draw[->, thin, color=midgray] (0,-0.40) -- (0, 0.80) node[above, font=\fontsize{6}{7}\selectfont] {$z_{t,\ell}$};
    \draw[dashed, thin, color=bordergray] (0,0) -- (4.9,0);

    % Shock impulses for Fed (blue)
    \draw[line width=1.3pt, color=navyblue] (1.6, 0) -- (1.6, 0.58) node[above, font=\fontsize{5.5}{6.5}\selectfont\bfseries] {$+z_{\text{Fed}}$};
    \draw[line width=0.8pt, color=navyblue] (2.3, 0) -- (2.3, -0.16);
    \draw[line width=0.8pt, color=navyblue] (0.9, 0) -- (0.9, 0.05);

    % Shock impulses for Oil (orange)
    \draw[line width=1.3pt, color=accentorange] (3.3, 0) -- (3.3, 0.54) node[above, font=\fontsize{5.5}{6.5}\selectfont\bfseries] {$+z_{\text{Oil}}$};
    \draw[line width=0.8pt, color=accentorange] (4.0, 0) -- (4.0, -0.20);
\end{scope}

% Unboxed label for shocks with comfortable padding above bottom line
\node[font=\fontsize{5.5}{6.5}\selectfont, color=midgray, anchor=north] at (13.0, 6.60) {
    Zero-mean narrative shocks $\mathbf{z}_t \in \mathbb{R}^{60}$
};

% =========================================================================
% CONNECTOR CORRIDOR: SHOCKS + STOCK RETURNS -> COVARIANCES
% Channel between y = 4.65 and y = 5.80 (height = 1.15)
% =========================================================================
% Shocks flow down and left
\draw[-, line width=1.1pt, color=navyblue] (13.0, 5.80) -- (13.0, 5.20) -- node[above, font=\fontsize{6}{7}\selectfont\bfseries, color=navyblue] {
    Narrative Shock Signals $\mathbf{z}_t \in \mathbb{R}^{60}$
} (2.65, 5.20);

% Stock returns enter from the left cleanly with text above arrow
\draw[->, line width=1.0pt, color=accentteal] (0.3, 5.20) -- node[above, font=\fontsize{5.8}{6.8}\selectfont\bfseries, color=accentteal] {Stock returns $r_{i,t}^e$} (2.55, 5.20);

% Junction dot
\fill[navyblue] (2.65, 5.20) circle (1.7pt);

% Joint downward feed into Stage 4
\draw[dataflow] (2.65, 5.20) -- (2.65, 4.65);

% =========================================================================
% BOTTOM PANEL: STAGE 4, 5 \& 6 (Covariances -> Sparse IPCA -> Factors & MVE)
% Dimensions: x from -0.1 to 16.1, y from -0.75 to 4.65 (height = 5.40)
% =========================================================================
\draw[panel] (-0.1, -0.75) rectangle (16.1, 4.65);
\node[anchor=north west, font=\bfseries\scriptsize\color{navyblue}] 
    at (3.2, 4.48) {\textsc{Stage 4, 5 \& 6: Cross-Sectional Pricing, Sparse IPCA \& Systematic Risk}};

% -------------------------------------------------------------------------
% 4. STOCK-NARRATIVE COVARIANCE EXPOSURES (x: 0.1 to 4.4)
% -------------------------------------------------------------------------
\node[header, anchor=north west] at (0.2, 4.05) {4. Stock Exposures};
\node[annot, anchor=north west] at (0.2, 3.72) {$\mathbf{c}_{i,t} \in \mathbb{R}^{61}$ (trailing 252d)};

% Heatmap matrix
\begin{scope}[shift={(1.0, 0.45)}]
    % Compact stock tickers strictly inside panel
    \node[font=\fontsize{5.8}{6.8}\selectfont\bfseries, anchor=east] at (-0.15, 1.95) {XOM};
    \node[font=\fontsize{5.8}{6.8}\selectfont\bfseries, anchor=east] at (-0.15, 1.35) {DAL};
    \node[font=\fontsize{5.8}{6.8}\selectfont\bfseries, anchor=east] at (-0.15, 0.75) {AAPL};
    \node[font=\fontsize{5.8}{6.8}\selectfont\bfseries, anchor=east] at (-0.15, 0.15) {JPM};

    % Simplified column headers
    \node[font=\fontsize{5.2}{6.2}\selectfont\bfseries, color=accentorange, rotate=40, anchor=south west] at (0.15, 2.3) {Oil};
    \node[font=\fontsize{5.2}{6.2}\selectfont\bfseries, color=navyblue, rotate=40, anchor=south west] at (0.75, 2.3) {Rates};
    \node[font=\fontsize{5.2}{6.2}\selectfont\bfseries, color=accentteal, rotate=40, anchor=south west] at (1.35, 2.3) {Tech};
    \node[font=\fontsize{5.2}{6.2}\selectfont, color=midgray, rotate=40, anchor=south west] at (1.95, 2.3) {$\dots \; \ell_{60}$};

    % Matrix cells: row 1 (XOM)
    \draw[fill=accentorange!75, draw=bordergray] (0.0, 1.65) rectangle (0.6, 2.25) node[pos=.5, font=\fontsize{5.5}{6.5}\selectfont\bfseries\color{white}] {+0.42};
    \draw[fill=softblue, draw=bordergray] (0.6, 1.65) rectangle (1.2, 2.25) node[pos=.5, font=\fontsize{5.5}{6.5}\selectfont] {+0.05};
    \draw[fill=lightgray, draw=bordergray] (1.2, 1.65) rectangle (1.8, 2.25) node[pos=.5, font=\fontsize{5.5}{6.5}\selectfont] {-0.02};
    \draw[fill=lightgray!60, draw=bordergray] (1.8, 1.65) rectangle (2.6, 2.25) node[pos=.5, font=\fontsize{5.5}{6.5}\selectfont] {$\dots$};

    % Matrix cells: row 2 (DAL)
    \draw[fill=navyblue!65, draw=bordergray] (0.0, 1.05) rectangle (0.6, 1.65) node[pos=.5, font=\fontsize{5.5}{6.5}\selectfont\bfseries\color{white}] {-0.38};
    \draw[fill=softorange, draw=bordergray] (0.6, 1.05) rectangle (1.2, 1.65) node[pos=.5, font=\fontsize{5.5}{6.5}\selectfont] {+0.08};
    \draw[fill=lightgray, draw=bordergray] (1.2, 1.05) rectangle (1.8, 1.65) node[pos=.5, font=\fontsize{5.5}{6.5}\selectfont] {+0.01};
    \draw[fill=lightgray!60, draw=bordergray] (1.8, 1.05) rectangle (2.6, 1.65) node[pos=.5, font=\fontsize{5.5}{6.5}\selectfont] {$\dots$};

    % Matrix cells: row 3 (AAPL)
    \draw[fill=lightgray, draw=bordergray] (0.0, 0.45) rectangle (0.6, 1.05) node[pos=.5, font=\fontsize{5.5}{6.5}\selectfont] {-0.03};
    \draw[fill=navyblue!40, draw=bordergray] (0.6, 0.45) rectangle (1.2, 1.05) node[pos=.5, font=\fontsize{5.5}{6.5}\selectfont\color{white}] {-0.18};
    \draw[fill=accentteal!75, draw=bordergray] (1.2, 0.45) rectangle (1.8, 1.05) node[pos=.5, font=\fontsize{5.5}{6.5}\selectfont\bfseries\color{white}] {+0.45};
    \draw[fill=lightgray!60, draw=bordergray] (1.8, 0.45) rectangle (2.6, 1.05) node[pos=.5, font=\fontsize{5.5}{6.5}\selectfont] {$\dots$};

    % Matrix cells: row 4 (JPM)
    \draw[fill=softorange, draw=bordergray] (0.0, -0.15) rectangle (0.6, 0.45) node[pos=.5, font=\fontsize{5.5}{6.5}\selectfont] {+0.06};
    \draw[fill=accentorange!70, draw=bordergray] (0.6, -0.15) rectangle (1.2, 0.45) node[pos=.5, font=\fontsize{5.5}{6.5}\selectfont\bfseries\color{white}] {+0.35};
    \draw[fill=lightgray, draw=bordergray] (1.2, -0.15) rectangle (1.8, 0.45) node[pos=.5, font=\fontsize{5.5}{6.5}\selectfont] {+0.04};
    \draw[fill=lightgray!60, draw=bordergray] (1.8, -0.15) rectangle (2.6, 0.45) node[pos=.5, font=\fontsize{5.5}{6.5}\selectfont] {$\dots$};
\end{scope}

% Unboxed label below matrix with comfortable margin to bottom border
\node[font=\fontsize{5.8}{6.8}\selectfont, color=navyblue, anchor=north] at (2.2, 0.05) {
    \textbf{Characteristic matrix $\mathbf{C}_t \in \mathbb{R}^{N_t \times 61}$}
};
\node[font=\fontsize{5.2}{6.2}\selectfont, color=midgray, anchor=north] at (2.2, -0.22) {
    Standardized, includes intercept
};

% Arrow 4 -> 5
\draw[dataflow] (4.0, 1.55) -- node[above, font=\fontsize{5.8}{6.8}\selectfont\bfseries, color=navyblue] {Exposures} 
                              node[below, font=\fontsize{5.2}{6.2}\selectfont, color=midgray] {$\mathbf{C}_t$} (5.0, 1.55);

% -------------------------------------------------------------------------
% 5. FACTOR EXTRACTION: SPARSE IPCA (x: 5.0 to 9.8)
% -------------------------------------------------------------------------
\node[header, anchor=north west] at (5.1, 4.05) {5. Sparse IPCA Extraction};
\node[annot, anchor=north west] at (5.1, 3.72) {Group-sparse narrative loadings};

% Matrix Gamma illustration
\begin{scope}[shift={(5.5, 1.10)}]
    % Gamma matrix frame (61 rows x 3 columns)
    \draw[line width=0.9pt, draw=navyblue] (0, -1.0) rectangle (1.3, 1.7);
    \node[above, font=\fontsize{6}{7}\selectfont\bfseries, color=navyblue] at (0.65, 1.7) {$\boldsymbol{\Gamma} \;(61 \times 3)$};
    \node[above, font=\fontsize{5.5}{6.5}\selectfont, color=midgray] at (0.22, 1.35) {$f_1$};
    \node[above, font=\fontsize{5.5}{6.5}\selectfont, color=midgray] at (0.65, 1.35) {$f_2$};
    \node[above, font=\fontsize{5.5}{6.5}\selectfont, color=midgray] at (1.08, 1.35) {$f_3$};

    % Active row 1: Fed (blue)
    \draw[fill=navyblue!35, draw=navyblue] (0, 1.05) rectangle (1.3, 1.35);
    \node[right, font=\fontsize{5.5}{6.5}\selectfont, anchor=west] at (1.35, 1.20) {
        \textbf{\color{navyblue}Fed} \ \color{midgray}active
    };

    % Active row 2: Oil (orange)
    \draw[fill=accentorange!35, draw=accentorange] (0, 0.60) rectangle (1.3, 0.90);
    \node[right, font=\fontsize{5.5}{6.5}\selectfont, anchor=west] at (1.35, 0.75) {
        \textbf{\color{accentorange}Oil} \ \color{midgray}active
    };

    % Inactive row (Zero / eliminated)
    \draw[fill=gray!15, draw=bordergray] (0, 0.15) rectangle (1.3, 0.45);
    \node[right, font=\fontsize{5.5}{6.5}\selectfont, anchor=west] at (1.35, 0.30) {
        \color{midgray}— \ zeroed
    };

    % Active row 3: Tech (teal)
    \draw[fill=accentteal!35, draw=accentteal] (0, -0.30) rectangle (1.3, 0.00);
    \node[right, font=\fontsize{5.5}{6.5}\selectfont, anchor=west] at (1.35, -0.15) {
        \textbf{\color{accentteal}Tech} \ \color{midgray}active
    };

    % Ellipsis rows
    \node[font=\fontsize{6}{7}\selectfont, color=midgray] at (0.65, -0.65) {$\vdots$};
\end{scope}

% Unboxed label below Gamma with comfortable margin to bottom border
\node[font=\fontsize{5.8}{6.8}\selectfont, color=navyblue, anchor=north] at (6.8, 0.05) {
    \textbf{Group-sparse selection ($\sum_\ell \|\boldsymbol{\Gamma}_\ell\|_2$)}
};
\node[font=\fontsize{5.2}{6.2}\selectfont, color=midgray, anchor=north] at (6.8, -0.22) {
    53 of 60 narratives retained in ST
};

% Arrow 5 -> 6
\draw[dataflow] (9.4, 1.55) -- node[above, font=\fontsize{5.8}{6.8}\selectfont\bfseries, color=navyblue] {$K=3$} 
                              node[below, font=\fontsize{5.2}{6.2}\selectfont, color=midgray] {Factors $\mathbf{f}_t$} (10.4, 1.55);

% -------------------------------------------------------------------------
% 6. ECONOMIC OUTPUT: LATENT FACTORS & MVE PORTFOLIO (x: 10.4 to 15.9)
% -------------------------------------------------------------------------
\node[header, anchor=north west] at (10.5, 4.05) {6. Systematic Narrative Risk};
\node[annot, anchor=north west] at (10.5, 3.72) {Out-of-sample MVE tangency portfolio};

% 3 Latent factor series & cumulative return
\begin{scope}[shift={(10.6, 1.40)}]
    \draw[->, thin, color=midgray] (0,0) -- (4.7, 0);
    % x-axis label anchored inside right edge strictly within panel
    \node[font=\fontsize{5.5}{6.5}\selectfont, color=midgray, anchor=north east] at (4.7, -0.06) {$t$ (OOS)};
    \draw[->, thin, color=midgray] (0,-0.65) -- (0, 1.45) node[above, font=\fontsize{6}{7}\selectfont] {$R_t$};

    % Latent factor traces (faint)
    \draw[line width=0.6pt, color=navyblue!40] plot[smooth, tension=0.7] coordinates {
        (0.2, -0.28) (1.0, 0.18) (1.8, -0.10) (2.8, 0.38) (3.8, 0.10) (4.5, 0.48)
    };
    \node[font=\fontsize{5}{6}\selectfont\bfseries, color=navyblue!70] at (2.8, 0.52) {$f_1$};
    
    \draw[line width=0.6pt, color=accentorange!40] plot[smooth, tension=0.7] coordinates {
        (0.2, 0.18) (1.0, -0.38) (2.0, 0.28) (3.0, -0.18) (4.0, 0.38) (4.5, 0.10)
    };
    \node[font=\fontsize{5}{6}\selectfont\bfseries, color=accentorange!70] at (4.0, 0.52) {$f_2$};
    
    \draw[line width=0.6pt, color=accentteal!40] plot[smooth, tension=0.7] coordinates {
        (0.2, -0.10) (1.2, 0.28) (2.2, -0.28) (3.2, 0.10) (4.1, -0.10) (4.5, 0.28)
    };
    \node[font=\fontsize{5}{6}\selectfont\bfseries, color=accentteal!70] at (1.2, 0.42) {$f_3$};

    % Cumulative MVE Return Curve (bold primary line)
    \draw[line width=1.4pt, color=navyblue] plot[smooth, tension=0.7] coordinates {
        (0.0, 0.0) (0.8, 0.18) (1.5, 0.40) (2.2, 0.32) (3.0, 0.62) (3.8, 0.82) (4.5, 1.10)
    };
    \node[font=\fontsize{5.8}{6.8}\selectfont\bfseries, color=navyblue, anchor=south east] at (4.5, 1.12) {MVE Portfolio};
    \node[font=\fontsize{5.2}{6.2}\selectfont, color=midgray] at (2.2, -0.44) {Tangency: $\mathbf{w} = \boldsymbol{\Sigma}_f^{-1}\boldsymbol{\mu}_f$};
\end{scope}

% Unboxed label below plot with comfortable margin to bottom border
\node[font=\fontsize{5.8}{6.8}\selectfont, color=navyblue, anchor=north] at (13.1, 0.05) {
    \textbf{3 latent factors $\longrightarrow$ OOS MVE portfolio}
};
\node[font=\fontsize{5.2}{6.2}\selectfont, color=midgray, anchor=north] at (13.1, -0.22) {
    Evaluated over September 2020--January 2024
};

\end{tikzpicture}

%% file: main.bbl
\begin{thebibliography}{37}
\providecommand{\natexlab}[1]{#1}
\providecommand{\url}[1]{\texttt{#1}}
\expandafter\ifx\csname urlstyle\endcsname\relax
  \providecommand{\doi}[1]{doi: #1}\else
  \providecommand{\doi}{doi: \begingroup \urlstyle{rm}\Url}\fi

\bibitem[Angelov(2020)]{angelov2020top2vec}
Dimo Angelov.
\newblock {Top2Vec}: Distributed representations of topics.
\newblock \emph{arXiv preprint arXiv:2008.09470}, 2020.

\bibitem[Bailey and L{\'o}pez~de Prado(2014)]{bailey2014deflated}
David~H. Bailey and Marcos L{\'o}pez~de Prado.
\newblock The deflated {S}harpe ratio: Correcting for selection bias, backtest
  overfitting, and non-normality.
\newblock \emph{The Journal of Portfolio Management}, 40\penalty0 (5):\penalty0
  94--107, 2014.
\newblock \doi{10.3905/jpm.2014.40.5.094}.

\bibitem[Banerjee et~al.(2005)Banerjee, Dhillon, Ghosh, and
  Sra]{banerjee2005vmf}
Arindam Banerjee, Inderjit~S. Dhillon, Joydeep Ghosh, and Suvrit Sra.
\newblock Clustering on the unit hypersphere using von {M}ises-{F}isher
  distributions.
\newblock \emph{Journal of Machine Learning Research}, 6:\penalty0 1345--1382,
  2005.

\bibitem[Bell et~al.(2025)Bell, Kakhbod, Nazemi, Stanton, and
  Wallace]{bell2025fairness}
Sebastian Bell, Ali Kakhbod, Abdolreza Nazemi, Richard Stanton, and Nancy
  Wallace.
\newblock Fairness by design: Machine learning and interpretable mortgage
  lending.
\newblock \emph{Available at SSRN 6221578}, 2025.

\bibitem[Bell et~al.(2026{\natexlab{a}})Bell, Kakhbod, Lettau, and
  Nazemi]{bell2026alphaglass}
Sebastian Bell, Ali Kakhbod, Martin Lettau, and Abdolreza Nazemi.
\newblock {AlphaGlass}: Interpretable characteristic-based portfolio choice.
\newblock Working paper, National Bureau of Economic Research,
  2026{\natexlab{a}}.

\bibitem[Bell et~al.(2026{\natexlab{b}})Bell, Kakhbod, Lettau, and
  Nazemi]{bell2026glassbox}
Sebastian Bell, Ali Kakhbod, Martin Lettau, and Abdolreza Nazemi.
\newblock Glass box machine learning and corporate bond returns.
\newblock \emph{Journal of Financial Economics}, 181:\penalty0 104294,
  2026{\natexlab{b}}.

\bibitem[Bianchi et~al.(2021)Bianchi, Terragni, and
  Hovy]{bianchi2021pretraining}
Federico Bianchi, Silvia Terragni, and Dirk Hovy.
\newblock Pre-training is a hot topic: Contextualized document embeddings
  improve topic coherence.
\newblock In \emph{Proceedings of ACL-IJCNLP (Short Papers)}, pages 759--766,
  2021.
\newblock \doi{10.18653/v1/2021.acl-short.96}.

\bibitem[Blei et~al.(2003)Blei, Ng, and Jordan]{blei2003lda}
David~M. Blei, Andrew~Y. Ng, and Michael~I. Jordan.
\newblock Latent {D}irichlet allocation.
\newblock \emph{Journal of Machine Learning Research}, 3:\penalty0 993--1022,
  2003.

\bibitem[Bybee et~al.(2023)Bybee, Kelly, and Su]{bybee2023narrative}
Leland Bybee, Bryan Kelly, and Yinan Su.
\newblock Narrative asset pricing: Interpretable systematic risk factors from
  news text.
\newblock \emph{The Review of Financial Studies}, 36\penalty0 (12):\penalty0
  4759--4787, 2023.

\bibitem[Bybee et~al.(2024)Bybee, Kelly, Manela, and Xiu]{bybee2024business}
Leland Bybee, Bryan Kelly, Asaf Manela, and Dacheng Xiu.
\newblock Business news and business cycles.
\newblock \emph{The Journal of Finance}, 79\penalty0 (5):\penalty0 3105--3147,
  2024.
\newblock \doi{10.1111/jofi.13377}.

\bibitem[{Center for Research in Security Prices}(n.d.)]{crspstockdatabase}
{Center for Research in Security Prices}.
\newblock {CRSP US Stock Databases}.
\newblock \url{https://www.crsp.org/research/crsp-us-stock-databases/}, n.d.
\newblock Daily stock data obtained through Wharton Research Data Services
  (WRDS).

\bibitem[Chen et~al.(2023)Chen, Kakhbod, Kazemi, and Xing]{chen2023process}
Hui Chen, Ali Kakhbod, Maziar Kazemi, and Hao Xing.
\newblock \emph{Process intangibles and agency conflicts}, volume 4351116.
\newblock SSRN, 2023.

\bibitem[Cong et~al.(2025)Cong, Liang, Zhang, and Zhu]{cong2025textual}
Lin~William Cong, Tengyuan Liang, Xiao Zhang, and Wu~Zhu.
\newblock Textual factors: A scalable, interpretable, and data-driven approach
  to analyzing unstructured information.
\newblock \emph{Management Science}, 71\penalty0 (12):\penalty0 10727--10739,
  2025.
\newblock \doi{10.1287/mnsc.2020.01180}.

\bibitem[Devlin et~al.(2019)Devlin, Chang, Lee, and Toutanova]{devlin2019bert}
Jacob Devlin, Ming-Wei Chang, Kenton Lee, and Kristina Toutanova.
\newblock {BERT}: Pre-training of deep bidirectional transformers for language
  understanding.
\newblock In \emph{Proceedings of NAACL-HLT}, pages 4171--4186, 2019.
\newblock \doi{10.18653/v1/N19-1423}.

\bibitem[Dhillon and Modha(2001)]{dhillon2001concept}
Inderjit~S. Dhillon and Dharmendra~S. Modha.
\newblock Concept decompositions for large sparse text data using clustering.
\newblock \emph{Machine Learning}, 42\penalty0 (1):\penalty0 143--175, 2001.
\newblock \doi{10.1023/A:1007612920971}.

\bibitem[Dieng et~al.(2020)Dieng, Ruiz, and Blei]{dieng2020etm}
Adji~B. Dieng, Francisco J.~R. Ruiz, and David~M. Blei.
\newblock Topic modeling in embedding spaces.
\newblock \emph{Transactions of the Association for Computational Linguistics},
  8:\penalty0 439--453, 2020.
\newblock \doi{10.1162/tacl_a_00325}.

\bibitem[Dong et~al.(2024)Dong, Fan, and Peng]{dong2024fnspid}
Zihan Dong, Xinyu Fan, and Zhiyuan Peng.
\newblock {FNSPID}: A comprehensive financial news dataset in time series.
\newblock \emph{arXiv preprint arXiv:2402.06698}, 2024.

\bibitem[Dyer et~al.(2017)Dyer, Lang, and Stice-Lawrence]{dyer2017evolution}
Travis Dyer, Mark Lang, and Lorien Stice-Lawrence.
\newblock The evolution of 10-{K} textual disclosure: Evidence from latent
  {D}irichlet allocation.
\newblock \emph{Journal of Accounting and Economics}, 64\penalty0
  (2--3):\penalty0 221--245, 2017.
\newblock \doi{10.1016/j.jacceco.2017.07.002}.

\bibitem[Fedyk et~al.(2026)Fedyk, Kakhbod, Li, and Malmendier]{fedyk2026ai}
Anastassia Fedyk, Ali Kakhbod, Peiyao Li, and Ulrike Malmendier.
\newblock {AI} and perception biases in investments: An experimental study.
\newblock \emph{Journal of Financial Economics}, 185:\penalty0 104350, 2026.

\bibitem[Gentzkow et~al.(2019)Gentzkow, Kelly, and Taddy]{gentzkow2019text}
Matthew Gentzkow, Bryan Kelly, and Matt Taddy.
\newblock Text as data.
\newblock \emph{Journal of Economic Literature}, 57\penalty0 (3):\penalty0
  535--574, 2019.
\newblock \doi{10.1257/jel.20181020}.

\bibitem[Gibbons et~al.(1989)Gibbons, Ross, and Shanken]{gibbons1989test}
Michael~R. Gibbons, Stephen~A. Ross, and Jay Shanken.
\newblock A test of the efficiency of a given portfolio.
\newblock \emph{Econometrica}, 57\penalty0 (5):\penalty0 1121--1152, 1989.
\newblock \doi{10.2307/1913625}.

\bibitem[Griffiths and Steyvers(2004)]{griffiths2004finding}
Thomas~L. Griffiths and Mark Steyvers.
\newblock Finding scientific topics.
\newblock \emph{Proceedings of the National Academy of Sciences}, 101\penalty0
  (suppl 1):\penalty0 5228--5235, 2004.
\newblock \doi{10.1073/pnas.0307752101}.

\bibitem[Grootendorst(2022)]{grootendorst2022bertopic}
Maarten Grootendorst.
\newblock {BERTopic}: Neural topic modeling with a class-based {TF-IDF}
  procedure.
\newblock \emph{arXiv preprint arXiv:2203.05794}, 2022.

\bibitem[Hochberg et~al.(2026)Hochberg, Kakhbod, Li, and
  Sachdeva]{hochberg2026patents}
Yael~V. Hochberg, Ali Kakhbod, Peiyao Li, and Kunal Sachdeva.
\newblock Are patents with female inventors under-cited? {E}vidence from text
  estimation.
\newblock \emph{Journal of Financial Economics}, 183:\penalty0 104307, 2026.
\newblock \doi{10.1016/j.jfineco.2026.104307}.

\bibitem[Hoyle et~al.(2021)Hoyle, Goel, Hian-Cheong, Peskov, Boyd-Graber, and
  Resnik]{hoyle2021broken}
Alexander Hoyle, Pranav Goel, Andrew Hian-Cheong, Denis Peskov, Jordan
  Boyd-Graber, and Philip Resnik.
\newblock Is automated topic model evaluation broken? {T}he incoherence of
  coherence.
\newblock In \emph{Advances in Neural Information Processing Systems},
  volume~34, pages 2018--2033, 2021.

\bibitem[Huang et~al.(2023)Huang, Wang, and Yang]{huang2023finbert}
Allen~H. Huang, Hui Wang, and Yi~Yang.
\newblock {FinBERT}: A large language model for extracting information from
  financial text.
\newblock \emph{Contemporary Accounting Research}, 40\penalty0 (2):\penalty0
  806--841, 2023.
\newblock \doi{10.1111/1911-3846.12832}.

\bibitem[Kakhbod and Li(2025)]{kakhbod2025nolbert}
Ali Kakhbod and Peiyao Li.
\newblock {NoLBERT}: A no lookahead(back) foundational language model for
  empirical research.
\newblock \emph{arXiv preprint arXiv:2509.01110}, 2025.

\bibitem[Kakhbod et~al.(2025{\natexlab{a}})Kakhbod, Kazempour, Livdan, and
  Sch{\"u}rhoff]{kakhbod2025finfluencers}
Ali Kakhbod, Seyed~Mohammad Kazempour, Dmitry Livdan, and Norman Sch{\"u}rhoff.
\newblock Finfluencers.
\newblock Discussion Paper 20204, Centre for Economic Policy Research,
  2025{\natexlab{a}}.

\bibitem[Kakhbod et~al.(2025{\natexlab{b}})Kakhbod, Kogan, Li, and
  Papanikolaou]{kakhbod2025creative}
Ali Kakhbod, Leonid Kogan, Peiyao Li, and Dimitris Papanikolaou.
\newblock Measuring creative destruction.
\newblock MIT Sloan Research Paper No.~7234-24, 2025{\natexlab{b}}.

\bibitem[Kakhbod et~al.(2026)Kakhbod, Kermani, and Maciel]{kakhbod2026fedmind}
Ali Kakhbod, Amir Kermani, and Bernardo Maciel.
\newblock In the {Fed}'s mind.
\newblock Working Paper 35016, National Bureau of Economic Research, 2026.

\bibitem[Ke et~al.(2019)Ke, Kelly, and Xiu]{ke2019predicting}
Zheng~Tracy Ke, Bryan~T. Kelly, and Dacheng Xiu.
\newblock Predicting returns with text data.
\newblock Working Paper 26186, National Bureau of Economic Research, 2019.

\bibitem[Kelly et~al.(2019)Kelly, Pruitt, and Su]{kelly2019characteristics}
Bryan~T. Kelly, Seth Pruitt, and Yinan Su.
\newblock Characteristics are covariances: A unified model of risk and return.
\newblock \emph{Journal of Financial Economics}, 134\penalty0 (3):\penalty0
  501--524, 2019.
\newblock \doi{10.1016/j.jfineco.2019.05.001}.

\bibitem[Lopez-Lira(2019)]{lopezlira2019risk}
Alejandro Lopez-Lira.
\newblock Risk factors that matter: Textual analysis of risk disclosures for
  the cross-section of returns.
\newblock SSRN Working Paper No.~3313663, 2019.

\bibitem[Pham et~al.(2024)Pham, Hoyle, Sun, Resnik, and
  Iyyer]{pham2024topicgpt}
Chau~Minh Pham, Alexander Hoyle, Simeng Sun, Philip Resnik, and Mohit Iyyer.
\newblock {TopicGPT}: A prompt-based topic modeling framework.
\newblock In \emph{Proceedings of NAACL-HLT}, pages 2956--2984, 2024.
\newblock \doi{10.18653/v1/2024.naacl-long.164}.

\bibitem[Reimers and Gurevych(2019)]{reimers2019sentencebert}
Nils Reimers and Iryna Gurevych.
\newblock Sentence-{BERT}: Sentence embeddings using {S}iamese {BERT}-networks.
\newblock In \emph{Proceedings of EMNLP-IJCNLP}, pages 3982--3992, 2019.
\newblock \doi{10.18653/v1/D19-1410}.

\bibitem[R{\"o}der et~al.(2015)R{\"o}der, Both, and
  Hinneburg]{roder2015coherence}
Michael R{\"o}der, Andreas Both, and Alexander Hinneburg.
\newblock Exploring the space of topic coherence measures.
\newblock In \emph{Proceedings of the Eighth ACM International Conference on
  Web Search and Data Mining}, pages 399--408, 2015.
\newblock \doi{10.1145/2684822.2685324}.

\bibitem[Sia et~al.(2020)Sia, Dalmia, and Mielke]{sia2020tired}
Suzanna Sia, Ayush Dalmia, and Sabrina~J. Mielke.
\newblock Tired of topic models? {C}lusters of pretrained word embeddings make
  for fast and good topics too!
\newblock In \emph{Proceedings of EMNLP}, pages 1728--1736, 2020.
\newblock \doi{10.18653/v1/2020.emnlp-main.135}.

\end{thebibliography}
